\documentclass[a4paper]{article}

\providecommand{\keywords}[1]{\noindent \textbf{\textit{Keywords---}} #1}

\usepackage{cite}
\usepackage{amsmath,amssymb,amsfonts,amsthm}
\usepackage{authblk}

\usepackage{graphicx}
\usepackage{tikz}
\usepackage{circuitikz}
\usepackage{xcolor}
\newcommand{\Frac}[2]{\displaystyle{\frac{#1}{#2}}}

\allowdisplaybreaks

\begin{document}

\title{Cardiovascular function and ballistocardiogram: a relationship interpreted via mathematical modeling}

\date{}
\author[1]{Giovanna Guidoboni}
\author[2]{Lorenzo Sala}
\author[3]{Moein Enayati}
\author[4]{Riccardo Sacco}
\author[5]{Marcela Szopos}
\author[3]{James Keller}
\author[6]{Mihail Popescu}
\author[7]{Laurel Despins}
\author[8]{Virginia H. Huxley}
\author[3]{Marjorie Skubic}

\affil[1]{\scriptsize Department of Electrical Engineering and Computer Science and with the Department of Mathematics, University of Missouri, Columbia, MO, 65211 USA \newline \emph{email: guidobonig@missouri.edu}.}
\affil[2]{\scriptsize Universit\'e de Strasbourg, CNRS, IRMA UMR 7501, Strasbourg, France.}
\affil[5]{\scriptsize Universit\'e Paris Descartes, MAP5, UMR CNRS 8145, Paris, France.}
\affil[3]{\scriptsize Department of Electrical Engineering and Computer Science, University of Missouri, Columbia, MO, 65211 USA.}
\affil[4]{\scriptsize Dipartimento di Matematica, Politecnico di Milano, Piazza Leonardo da Vinci 32, 20133 Milano, Italy.}
\affil[6]{\scriptsize Department of Health Management and Informatics, University of Missouri, Columbia, MO, 65211 USA.}
\affil[7]{\scriptsize Sinclair School of Nursing, University of Missouri, Columbia, MO, 65211 USA.}
\affil[8]{\scriptsize Department of Medical Pharmacology and Physiology, University of Missouri, Columbia, MO, 65211 USA.}

\maketitle

{\let\thefootnote\relax\footnote{{\footnotesize Copyright (c) 2017 IEEE. Personal use of this material is permitted. However, permission to use this material for any other purposes must be obtained from the IEEE by sending an email to pubs-permissions@ieee.org.}}}

\vspace{-15mm}

\begin{abstract} 
\noindent
\emph{Objective:} to develop quantitative methods for the clinical interpretation of the ballistocardiogram (BCG). \\
\emph{Methods:} a closed-loop mathematical model of the cardiovascular system is proposed to theoretically simulate the mechanisms generating the BCG signal, which is then compared with the signal acquired via accelerometry on a suspended bed. \\
\emph{Results:} simulated arterial pressure waveforms and ventricular functions are in good qualitative and quantitative agreement with those reported in the clinical literature. Simulated BCG signals exhibit the typical I, J, K, L, M and N peaks and show good qualitative and quantitative agreement with experimental measurements. Simulated BCG signals associated with reduced contractility and increased stiffness of the left ventricle exhibit different changes that are characteristic of the specific pathological condition. \\
\emph{Conclusion:} the proposed closed-loop model captures the predominant features of BCG signals {and can predict pathological changes} on the basis of fundamental mechanisms in cardiovascular physiology. \\
\emph{Significance:} this work provides a quantitative framework for the clinical interpretation of BCG signals {and the optimization of BCG sensing devices}. The present study considers {an average human body and can potentially be extended to include variability among individuals}. 
\end{abstract}


\maketitle

\section{Introduction}
\label{sec:intro}
Ballistocardiography captures the signal generated by the repetitive motion of the human body due to sudden ejection of blood into the great vessels with each heartbeat~\cite{StarrNoordergraafBook}. The signal is called ballistocardiogram (BCG). Extensive research work by Starr and Noordergraaf showed that the effect of main heart malfunctions, such as congestive heart failure and valvular disease, would alter the BCG signal~\cite{StarrNoordergraafBook,Pinheiro2010,starr1961twenty}, 
thereby yielding a great potential for passive, noncontact monitoring of the cardiovascular status. 

The original measurement device used by Starr and others was a lightweight bed suspended by long cables. The blood flow of a subject lying on the suspended bed resulted in the bed swinging; the capture of the swing was the BCG signal. This measurement device was impractical for standardized BCG measurements, especially compared to the electrocardiogram (ECG), which could be taken on virtually any platform using electric leads placed on the body in a standard configuration. The standardization of the ECG measurement has allowed clinical interpretation of the ECG waveform, such that it can be used to diagnose abnormalities in cardiac function. 

Recently, there has been a resurgence of BCG research, as new sensing devices (e.g., in the form of bed sensors) allow easier, noninvasive capture of the BCG signal. In addition, the BCG offers advantages over the standardized ECG measurement, in that direct body contact is not required and the signal reflects the status of the larger cardiovascular system. These advantages provide intriguing possibilities of continuous passive monitoring of the cardiovascular system without requiring the patient to do anything. 

There have been a variety of bed, chair, and other sensors proposed to capture the BCG; several are now available commercially~\cite{inan2015ballistocardiography,shin2008automatic,inan2009robust,giovangrandi2011ballistocardiography,rate2007emfit,alametsa2008ballistocardiography,
paalasmaa2015adaptive,
chen2008unconstrained,
katz2016contact,
rosales2017heart,
huffaker2018passive,
young2008method,
zimlichman2012early,
helfand2016technology}. 
Much of this work has focused on monitoring heart rate along with respiration rate from the accompanying respiration signal, and other parameters for tracking sleep quality. Recent work has also investigated the BCG waveform morphology for the purpose of tracking changes in cardiovascular health~\cite{su2018monitoring,Pinheiro2010,javaid2016elucidating}.
 This offers a special relevance and significant potential in monitoring older adults as they age. Identifying very early signs of cardiovascular health changes provides an opportunity for very early treatments before health problems escalate; very early treatment offers better health outcomes and the potential to avoid hospitalizations~\cite{rantz2015enhanced,rantz2015new}. 
 
 One challenge in using the BCG waveform to track cardiovascular health changes is the lack of a standardized measurement device and protocol and, thus, the lack of uniform clinical interpretation of the BCG signal across the various sensing devices. Here, we offer a first step in addressing this challenge by building a theoretical BCG signal based on a novel closed-loop mathematical model of the cardiovascular system and  {predicting how the BCG signal would change in the presence of specific pathological conditions.}
 Parameters of the model are calibrated on known physiological conditions for the average human body (see Appendix~\ref{app:model}). 
 Subsequent work will include models for different BCG sensing systems. Thus, BCG signals captured by different sensing systems could be translated into a standard BCG signal for uniform clinical interpretation. 
 
The paper is organized as follows. In Section~\ref{sec:background}, we review the background and related work. 
Section~\ref{sec:methods} describes the proposed mathematical model and our validation approach with a replica of Starr's suspended bed built within the Center for Eldercare and Rehabilitation Technology at the University of Missouri (see Fig.~\ref{fig:hanging_bed}). 
Further details on the mathematical model are included in Appendix~\ref{app:model}. 
In Section~\ref{sec:results}, results are presented which compare the output of the proposed model to experimental waveforms from related work, as well as our own validation with a replica of the suspended bed.  {BCG signals associated with reduced contractility and increased stiffness of the left ventricle are also simulated and compared.}
Conclusions and future research directions are summarized in Section~\ref{sec:conclusions}. 

%
\begin{figure}[h!]
\centering
\includegraphics[width=0.85\textwidth]{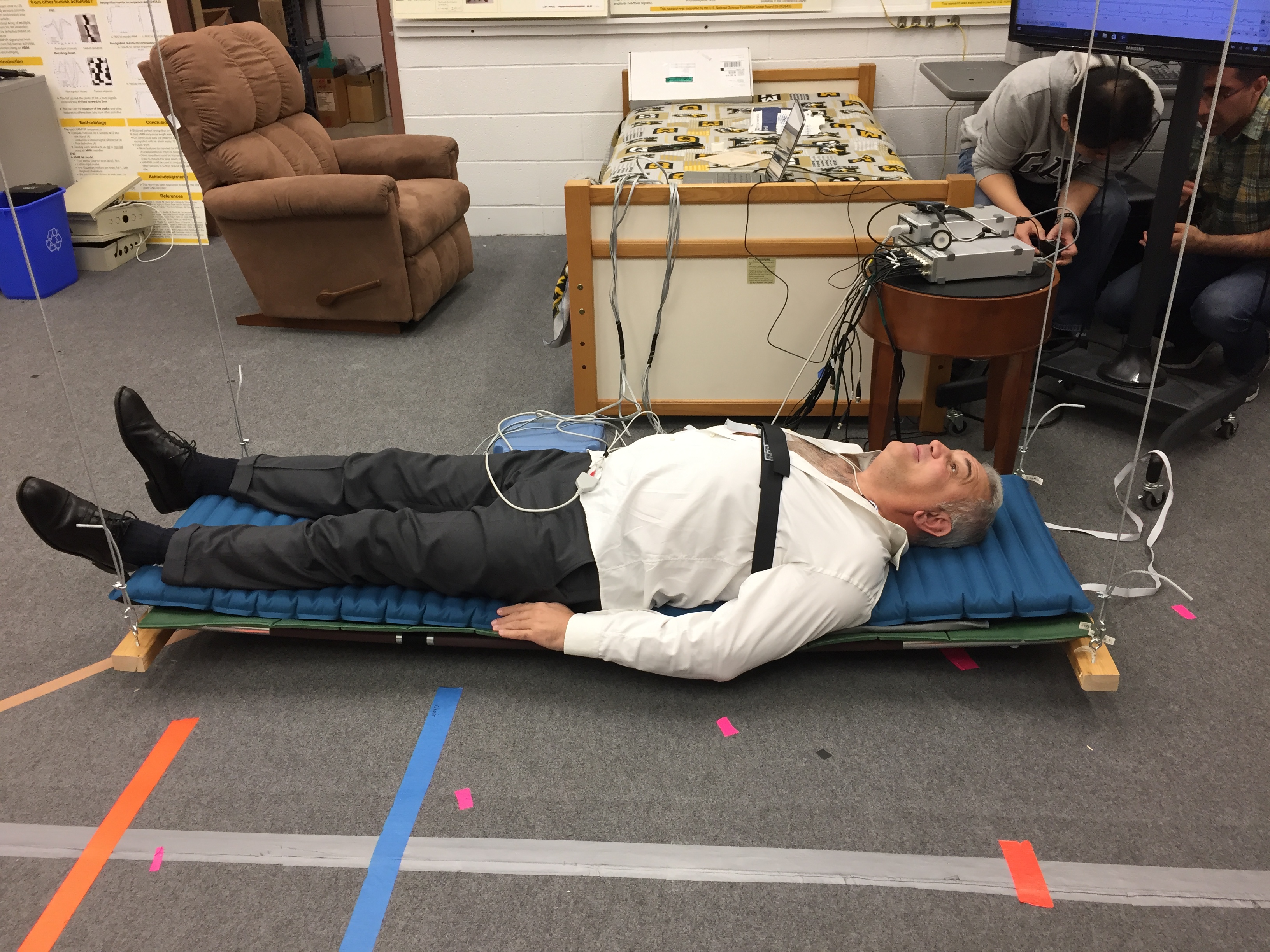}
\caption{Replica of the original Starr's suspended bed built within the Center for Eldercare and Rehabilitation Technology at the University of Missouri.
}
\label{fig:hanging_bed}
\end{figure}


\section{Background and related Work}
\label{sec:background}

 {Starr} and Noordergraaf provided the theoretical foundations to interpret BCG signals by expressing the displacement of the center of mass of the human body as a function of the blood volumes occupying different vascular compartments at a given time during the cardiac cycle~\cite{StarrNoordergraafBook}. Specifically,
the coordinate $Y$ of the center of mass of the body along the head-to-toe direction at any given time $t$ can be written as
\begin{equation}\label{eq:BCG}
Y(t) = \frac{\rho_b}{M}  \sum_{i=1}^N V_i(t) y_i + c\,,
\end{equation}
where $\rho_b$ is the blood density, $M$ is the body mass, $N$ is the total number of vascular compartments 
in the model and $c$ is a constant term representing the body frame. Each vascular compartment $i$, with $i=1,...,N$, is assumed to be located at the fixed coordinate $y_i$ and to be filled with the blood volume $V_i(t)$ at time $t$. 
Since the term $c$ in Eq.~(\ref{eq:BCG}) is constant for a given person,  the BCG signal associated with the center of mass displacement in the head-to-toe direction is defined as
\begin{equation}\label{eq:BCG_disp}
\mbox{BCG}_{disp}(t) := \frac{\rho_b}{M}  \sum_{i=1}^N V_i(t) y_i \,.
\end{equation}
The BCG signals associated with velocity and acceleration of the center of mass can be obtained via time differentiation as 
\begin{equation}\label{eq:BCG_vel}
\mbox{BCG}_{vel}(t) := \frac{\rho_b}{M}  \sum_{i=1}^N \frac{dV_i(t)}{dt} y_i \,,
\end{equation}
\begin{equation}\label{eq:BCG_acc}
\mbox{BCG}_{acc}(t) := \frac{\rho_b}{M}  \sum_{i=1}^N \frac{d^2V_i(t)}{dt^2} y_i \,,
\end{equation}
respectively.  
In our body, the waveforms $V_i(t)$ result from the complex interplay between the blood volume ejected from the heart, the resistance to flow that blood experiences across the cardiovascular system and the pressure distribution within it. 
 {Starr} and Noordergraaf characterized the volume waveforms $V_i(t)$ by means of experimental measurements at each location $y_i$~\cite{StarrNoordergraafBook}. 
In the present work, the waveforms $V_i(t)$ are calculated by means of
a mathematical model based on the physical principles governing vascular physiology, thereby paving the way to the use of
quantitative methods to interpret BCG signals and identify cardiovascular abnormalities in a given patient. 

The first computer-aided approach for quantitative interpretation of BCG signals was proposed by Noordergraaf et al in~\cite{Noordergraaf1963}, where
the electric analogy to fluid flow was leveraged to describe the motion of blood through the arterial system during the cardiac cycle and calculate the resulting BCG signal. 
%
Since then, only a few studies have been directed to the theoretical  {construction and} interpretation of BCG signals.
In~\cite{Wiard2009}, Wiard et al utilized a three-dimensional finite element model for blood flow in the thoracic aorta to show that the traction at the vessel wall appears of similar magnitude to recorded BCG forces. 
In~\cite{Kim2016}, Kim et al proposed a simplified model based on the equilibrium forces within the aorta to show that blood pressure gradients in the ascending and descending aorta are major contributors to the BCG signal.

Despite their different approaches to blood flow modeling, the aforementioned studies share the common feature of focusing only on the arterial side of the cardiovascular system, thereby leading to \textit{open-loop} models of the circulation. In reality, though, our blood circulates within a  \textit{closed-loop} system and, as a consequence, hemodynamic changes observed at the level of the major arteries might be the result of changes occurring elsewhere within the closed-loop system. %
For example, left ventricular heart failure leads to an increase in fluid pressure that is transferred back to the lungs, ultimately damaging the right side of the heart and causing right heart failure~\cite{bogaard2009right}. %
Another example is given by venothromboembolism, a disorder manifested by deep vein thrombosis and pulmonary embolism. Deep vein thrombosis occurs when a blood clot forms in a vein, most often in the leg, but they can also form in the deep veins of the arm, splanchnic veins, and cerebral veins~\cite{di2016deep}. A pulmonary embolism occurs when a clot breaks loose and travels to the pulmonary circulation, causing thrombotic outflow obstruction and a sudden strain on the right ventricle~\cite{jolly2018pulmonary}. Sequelae of such an event include decreased right ventricular cardiac output, poor overall cardiac output, and decreased systemic blood pressure. 

In order to account for these important feedback mechanisms within our circulatory system, the present work aims at providing, for the first time, a  quantitative interpretation of the BCG signal by means of a
 \textit{closed-loop} mathematical model for 
the  cardiovascular system. Several modeling approaches have been proposed in cardiovascular research, see e.g.~\cite{kerckhoffs2007coupling,smith2004minimal,vzavcek1996numerical}. In particular, some works have highlighted the non-negligible effect of the feedback wave within a closed-loop circuit, see e.g. \cite{liang2005closed,olansen2000closed,blanco2013dimensionally,hirschvogel2017monolithic}.
We have leveraged the knowledge available in the modeling literature to develop a novel closed-loop model for the cardiovascular system that includes sufficient elements to reproduce  {theoretically} the BCG signal  {and predict changes in the signal associated with specific pathological conditions.
As a result, the proposed model holds promise to serve as a virtual laboratory where cardiovascular dysfunctions can be simulated and their manifestation on the BCG signal can be characterized. 
}


\section{Methods}
\label{sec:methods}
In the next two sections we describe the closed-loop model for the  {construction and} interpretation of BCG signals (Section~\ref{sec:math_model}), and the experimental methods employed to acquire BCG signals in our laboratory (Section~\ref{sec:exp_data}).  


\subsection{Mathematical model}
\label{sec:math_model}

\begin{figure}
\centering
\begin{tikzpicture}
\node[inner sep=0pt] at (0,0)
    {\includegraphics[width=0.85\textwidth]{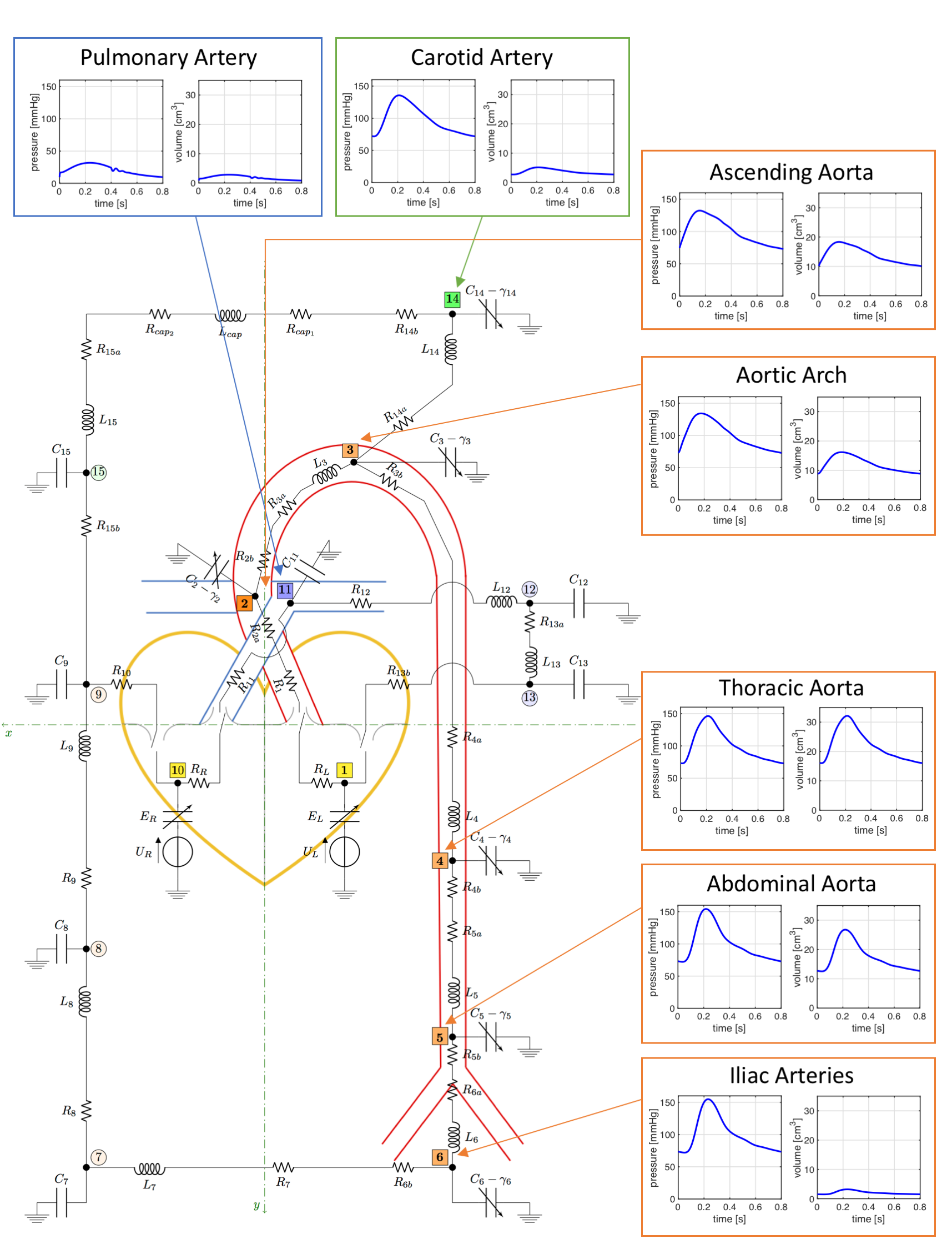}};    
\end{tikzpicture}
\caption{ \small 
	Schematic representation of the closed-loop model describing the flow of blood through the circulatory system. 
The model leverages the electric analogy to fluid flow, where resistors, capacitors and inductors represent hydraulic resistance, compliance and inertial effects, respectively. 
Arrows represent variable elements. The model comprises 4 main interconnected compartments: 
\emph{(i) the heart}, where the valves are represented by ideal switches and the ventricular pumps are modeled by means of voltage sources;
\emph{(ii) the systemic circulation}, where the arterial segments from the ascending aorta to the iliac arteries are modeled in great detail because of their relevance to the BCG signal; 
\emph{(iii) the pulmonary circulation}, which starts with the pulmonary artery at the right ventricle and converges into the left atrium; and
\emph{(iv) the cerebral circulation}, which branches from the aortic arch and converges into the systemic venous system. 
The circuit nodes have been marked by thick black dots and numeric labels from 1 to 14 highlighted in different colors (yellow, orange, blue, green) to distinguish among the 4 circulatory compartments included in the model (heart, systemic circulation, pulmonary circulation, cerebral circulation, respectively). Numerical labels in fuller color tones and square frames identify the circuit nodes utilized for the computation of the BCG signal. Arterial pressure and volume waveforms simulated via the closed-loop model are also included.
}
\label{fig:model}
\end{figure}

Blood circulation is modeled using the analogy between electric systems and hydraulic networks.
In this context, electric potentials correspond to fluid pressure, electric charges correspond to fluid volumes and electric currents correspond to volumetric flow rates. 
The closed-loop model  developed in this work consists of a network of resistors, capacitors, inductors, voltage sources and switches arranged into 4 main interconnected compartments representing the heart, the systemic circulation, the pulmonary circulation and the cerebral circulation, as  reported in Fig.~\ref{fig:model}. The circuit nodes have been marked by thick black dots and numeric labels from 1 to 14 highlighted in different colors (yellow, orange, blue, green) to distinguish among the 4 circulatory compartments included in the model (heart, systemic circulation, pulmonary circulation, cerebral circulation, respectively). The anatomical meaning of the circuit nodes has been summarized in Table~\ref{tab:yvalues}. 
In the model, resistors, capacitors and inductors represent hydraulic resistance, wall compliance and inertial effects, respectively. Variable capacitors, indicated with arrows in Fig.~\ref{fig:model}, are utilized to describe the wall viscoelasticity in large arteries and the nonlinear properties of the heart muscle fibers. Cerebral capillaries are assumed to be non-compliant as in~\cite{lakin2003whole}. 
The four heart valves are modeled as ideal switches. The ventricular pumps are modeled as time-dependent voltage sources. Ultimately, the model leads to a system of ordinary differential equations whose solution provides the time-dependent  profiles of pressures and volumes at the circuit nodes and flow rates through the circuit branches. A detailed description of the mathematical model and the related  model parameters is given in Appendix~\ref{app:model}.

\begin{table}[ht]
\caption{Anatomical meaning of the circuit nodes}
\centering
\begin{tabular}{llll}
\hline
\textsc{Node} & \textsc{Compartment} & \textsc{Segment} & \textsc{$y$} [cm] \\
\hline
1 & Heart & Left Ventricle &   0.5\\
2 & Systemic  &Ascending Aorta &  -2 \\
3 & Systemic  &Aortic Arch &   -7\\
4 & Systemic  &Thoracic Aorta &  20 \\
5 & Systemic  &Abdominal Aorta &  35 \\
6 & Systemic  &Iliac arteries &  45 \\
7 & Systemic & Small arteries & - \\
8 & Systemic & Capillaries & - \\
9 & Systemic & Veins & - \\
10 & Heart & Right Ventricle &  0.5 \\
11 & Pulmonary & Pulmonary Arteries & -5  \\
12 & Pulmonary & Capillaries &  - \\
13 & Pulmonary & Veins & -  \\
14& Cerebral & Cerebral Arteries &  -10 \\
15& Cerebral & Veins & -  \\
\hline 
\end{tabular}
\label{tab:yvalues}
\end{table}

The closed-loop model is utilized to  {construct 
the BCG signal theoretically} by substituting in Eqs.~(\ref{eq:BCG_disp})-(\ref{eq:BCG_acc}) the volume waveforms computed via the solution of the mathematical model. Precisely, we 
included 9 contributions
in our calculations for the BCG signal, meaning that $N=9$, which account for the volume waveforms pertaining to the left and right ventricles (nodes 1 and 10), four aortic segments (nodes 2, 3, 4, 5), iliac arteries (node 6), pulmonary arteries (node 11) and cerebral arteries (node 14). 
The circuit nodes whose volume waveforms are included in the BCG are indicated with fuller colors and square frames in Fig.~\ref{fig:model}. 
The calculation of the BCG signal requires the values of the $y-$coordinate representing the distance of each compartment of interest from an ideal plane through the heart, as depicted in Fig.~\ref{fig:model}. The values adopted in this work are summarized in Table~\ref{tab:yvalues}. 
The mathematical model has been implemented in OpenModelica~\cite{fritzson2005openmodelica}, an open-source Modelica-based modeling and simulation environment intended for industrial and academic studies of  complex dynamic systems.
Model results have been obtained using a differential algebraic system solver,
DASSL~\cite{petzold1982dassl}, with a tolerance of $10^{-6}$ and a time step of $0.001$~s.
We simulated $8$ cardiac cycles,  {each lasting 0.8~s}, for a total simulation time of  {$6.4$~s}, in order to obtain a periodic solution. 
Simulation results were post-processed using Matlab~\cite{matlab}, a commercial software to analyze data, develop algorithms and implement mathematical models. 
The results reported in Section~\ref{sec:results} correspond to the last of the $8$ simulated cardiac cycles.

\subsection{Experimental Methods}
\label{sec:exp_data}

A replica of the original suspended bed used by Starr, 
shown in Fig.~\ref{fig:hanging_bed},
was built within the Center for Eldercare and Rehabilitation Technology at the University of Missouri
 and was equipped with a three-axis accelerometer from Kionix Inc~\cite{Accelerometer_Datasheet}  with sensitivity of 1000~mV/g within the range of $\pm$2g. 
One healthy male subject of age 33, weight 78~kg and height 178~cm,  {lay on the bed} in the supine position for approximately 10 minutes. A  {pulse transducer from ADInstruments}
was wrapped around the ring finger of the subject's left hand;  {the pulse transducer detects expansion and contraction of the finger circumference due to changes in blood pressure. The signals from the three-axis accelerometer and the pulse transducer} have been collected simultaneously using an ADInstruments PowerLab Data Acquisition system \cite{ADI} at the rate of 1000 samples per second. 

The data acquired on the subject were processed in order to extract a representative BCG waveform,  {or template}, over a cardiac cycle.  {Due to the importance that specific choices for segmentation, filtering and alignment have on the resulting BCG template, we proceed by providing the details of our signal processing method}.
Local peaks of the pulse waveforms have been used as the reference for heartbeat activities, and segmentation of the BCG waveforms. 
Beats of length larger than 1.5~s or less than 0.4~s, corresponding to 40 and 150 beats per minute (bpm), respectively, were considered abnormal and were removed in the preprocessing.
As mentioned in Section~\ref{sec:intro}, the main focus of this study is
the $y-$axis signal of the accelerometer  since it provides the experimental waveform of $\mbox{BCG}_{acc} (t)$ defined in Eq.~(\ref{eq:BCG_acc}).  A band-pass Butterworth filter with cut-off frequencies at 0.7~Hz to 15~Hz was applied to eliminate low frequency respiratory base line and high frequency noises. The resulting acceleration signal has been segmented based on the timings previously acquired from the pulse  {transducer}.
Since the heart rate is naturally subject to variations, the length of each BCG cycle was not constant over the approximately 10 minutes of data acquisition. Thus, after removal of the outliers, the signals were cut and re-sampled to the median length of all, which resulted in 83~bpm for the subject under consideration. 
 
The work by Starr and Noordergraaf in \cite{StarrNoordergraafBook} at page 52, Fig. 4.3, was based on a cardiac cycle of 0.8~s, corresponding to 75~bpm. For the ease of comparison, we re-sampled our data to match the length of 0.8~s for the cardiac cycle. 
In order to further remove potential confounding effects due to respiratory movements, the mean was subtracted from each waveform, thereby making it zero-mean. 
Then, all waveforms were aligned to the median one, based on their cross-correlation value. 
All waveforms with correlation below 0.4 and lag-time above 0.4~s were considered to be motion artifacts and removed from the analysis.
The signals for velocity and displacement BCGs were obtained by integration starting from the  acceleration waveform. 

\section{Results}
\label{sec:results}
The capability of the proposed closed-loop model to capture some of the main features of the cardiovascular system is assessed in Section~\ref{sec:results_physio}. The comparison between BCG signals predicted theoretically via the closed-loop model and measured experimentally via the accelerometer on the suspended bed is presented in Section~\ref{sec:results_BCG}.  {Simulated BCG signals in the case of reduced contractility and increased stiffness of the left ventricle are compared in Section~\ref{sec:pathological_conditions}}.
\subsection{Cardiovascular physiology}
\label{sec:results_physio}

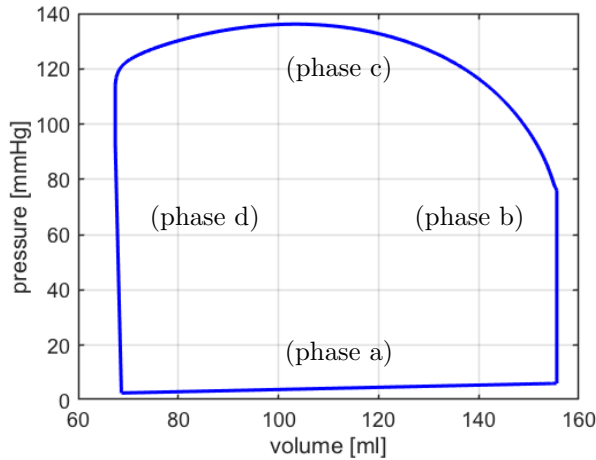
\begin{figure}
\centering
\begin{tikzpicture}[scale=0.95]
\node[inner sep=0pt] at (0,0)
    {\includegraphics[width=0.7\textwidth]{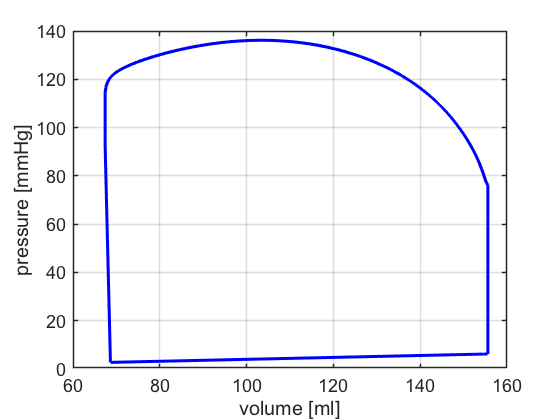}};
\node[above] at (0.3,-2.2) {(phase a)};
\node[left] at (3.,0) {(phase b)};
\node[below] at (0.3,2.4) {(phase c)};
\node[right] at (-2.45,0) {(phase d)};
\end{tikzpicture}
\caption{Volume-pressure relationship in the left ventricle during one cardiac cycle simulated via the closed-loop model. 
The model 
captures the four 
phases of ventricular function: ventricular filling (phase a), isovolumetric contraction (phase b), ejection (phase c), and isovolumetric relaxation (phase d)~\cite{klabunde2011cardiovascular}. 
}
\label{fig:VLV_PLV}
\end{figure}

A qualitative description of left ventricular function is shown in Fig.~\ref{fig:VLV_PLV}, which reports the volume-pressure relationship in the left ventricle during one cardiac cycle simulated via the closed-loop model. Results show that the 
model correctly captures the four basic phases of ventricular function: ventricular filling (phase a), isovolumetric contraction (phase b), ejection (phase c), and isovolumetric relaxation (phase d). 
%

\begin{figure}[h!]
\centering
\begin{tikzpicture}[scale=0.9]
\node[inner sep=0pt] at (0,-10)
    {\includegraphics[width=0.83\textwidth]{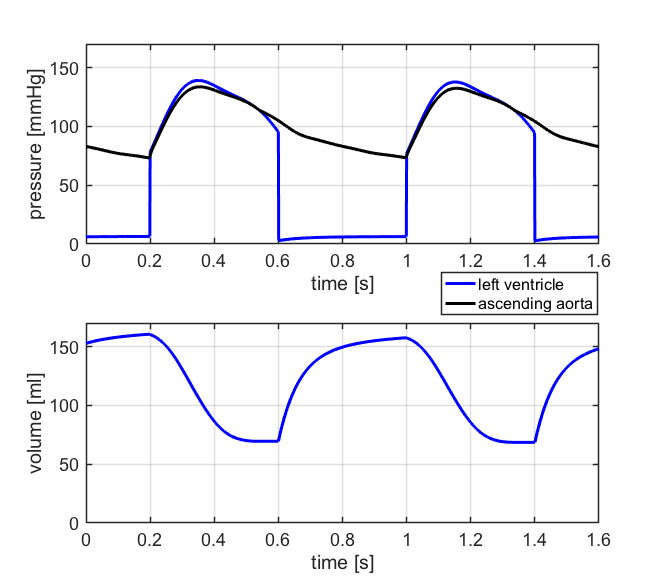}};
\end{tikzpicture}
\caption{Wiggers diagram simulated via the closed-loop model. Results show the typical features of isovolumetric contraction and relaxation~\cite{klabunde2011cardiovascular}.
}
\label{fig:wiggers_diagram}
\end{figure}

Another qualitative representation of ventricular function is given by the Wiggers diagram, where the volume waveform 
of the left ventricle is portrayed together with the pressure waveforms 
of the left ventricle and the ascending aorta. The Wiggers diagram simulated via the closed-loop model (see Fig.~\ref{fig:wiggers_diagram}) shows the typical features of isovolumetric contraction and relaxation exhibited by physiological waveforms~\cite{klabunde2011cardiovascular}.

%



Quantitative parameters describing cardiovascular physiology include end-diastolic volume (EDV), end-systolic volume (ESV), stroke volume (SV), cardiac output (CO) and ejection fraction (EF) associated with the left and right ventricles. EDV and ESV are computed as the maximum and minimum values of the ventricular volumes during the cardiac cycle, respectively, and their difference gives SV, namely SV = EDV-ESV. The relative difference between EDV and ESV gives EF, namely EF = 100 $\times$ (EDV-ESV)/EDV, which can also be written as EF = 100 $\times$ SV/EDV.
Then, denoting by $T_c$ the length of the heartbeat measured in seconds, the heart rate HR and the cardiac output CO are computed as HR = 60/$T_c$ and CO = HR $\times$ SV/1000.
Table~\ref{tab:Heart_Values} provides the values of these parameters for the left and right ventricles  as reported in the clinical literature and simulated via the closed-loop model. Specifically, the clinical studies in~\cite{Maceira2006Left} and~\cite{Maceira2006Right} utilized cardiovascular magnetic resonance to assess left and right ventricular functions on 120 healthy individuals.
All the simulated values fall within the ranges reported in the clinical literature, thereby validating the capability of the closed-loop model to capture the main features of the heart function.  

\begin{table}[h!]
\caption{Cardiovascular Physiology: Left and Right Ventricles}
\centering
\resizebox{\textwidth}{!}{
\begin{tabular}{llcccc}
\hline
\textsc{Parameter} & \textsc{Unit} &  \multicolumn{2}{c}{\textsc{Normal Clinical Range}} & \multicolumn{2}{c}{\textsc{Closed-Loop Model} } \\[.03in]
&& \emph{Left} & \emph{Right} & \emph{Left } & \emph{Right} \\
&& \emph{Ventricle} & \emph{Ventricle} & \emph{Ventricle} & \emph{Ventricle} \\
\hline\\
End-Diastolic  & [ml] & 
142 (102,183)~\cite{Maceira2006Left} &
144 (98,190)~\cite{Maceira2006Right} &
155.6 &
157.7 \\
Volume (EDV)& & & 100 - 160~\cite{Chart} 
\\[.05in]
End-Systolic  & [ml] & 
47 (27,68)~\cite{Maceira2006Left} &
50 (22,78)~\cite{Maceira2006Right} &
67.3 &
68.8\\
Volume (ESV)& & & 50-100~\cite{Chart} 
 \\[.05in]
Stroke  & [ml/beat] & 
95 (67, 123)~\cite{Maceira2006Left} &
94 (64, 124)~\cite{Maceira2006Right} &
88.2&
88.8\\
Volume (SV)& & 60 - 100~\cite{Chart}& 60-100~\cite{Chart} 
 \\[.05in]
Cardiac & [l/min] & 
4-8~\cite{Chart}  &
4-8~\cite{Chart} &
6.6&
6.6\\
Output (CO) & & & 
 \\[.05in]
Ejection & [\%] & 
67 (58, 76)~\cite{Maceira2006Left}   &
66 (54, 78)~\cite{Maceira2006Right}&
56.7&
56.3\\
Fraction (EF) & & & 40 - 60~\cite{Chart} \\[.05in]
\hline \\[.08in]
\end{tabular}}
\label{tab:Heart_Values}
\end{table}
%

\begin{figure}[h!]
\centering
\includegraphics[width=0.85\textwidth]{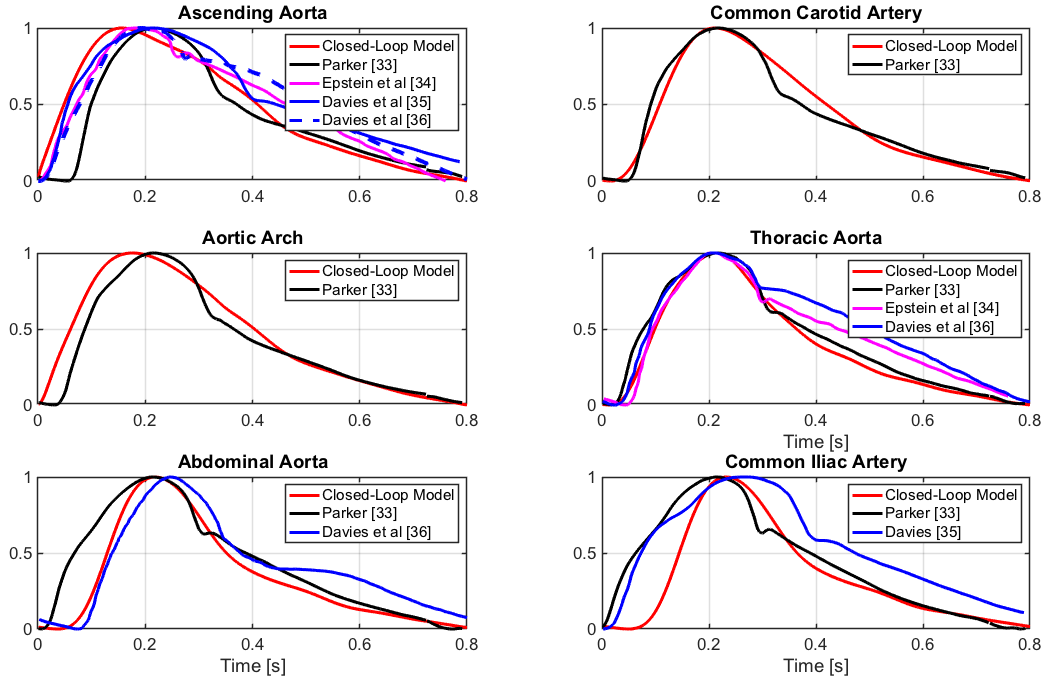}
\caption{Pressure waveforms simulated via the 
closed-loop model (\emph{red}) are compared with \emph{(i)} experimental measurements 
reported by Parker~\cite{parker2009introduction} (\emph{black}) and Davies et al~\cite{davies2012attenuation,davies2007importance} (\emph{blue}); and \emph{(ii)} theoretical simulations generated via a one-dimensional arterial network by Epstein et al~\cite{epstein2015reducing} (\emph{magenta}). 
}
\label{fig:comparison_arterial_pressure}
\end{figure}

The pressure and volume waveforms pertaining to the main segments of the systemic arteries simulated via the closed-loop model are reported in Fig.~\ref{fig:model}. Results show that the model captures the typical features of peak magnification and time delay exhibited by physiological waveforms~\cite{mcdonaldblood}. In Fig.~\ref{fig:comparison_arterial_pressure},
the simulated pressure waveforms are compared with experimental measurements at different sites along the arterial tree. 
In particular, we compare the results of our simulations with those presented by:
\emph{(i)} Davies et al~\cite{davies2012attenuation,davies2007importance}, where sensor-tipped intra-arterial wires are used to measure pressure in nearly 20 subjects (age, 35-73 years) at 10~cm intervals along the aorta, starting at the aortic root;
 \emph{(ii)} Parker~\cite{parker2009introduction}, where aortic pressures at 10, 20, 30, 40, 50 and 60~cm downstream from the aortic valve are reported; and 
  \emph{(iii)} Epstein et al~\cite{epstein2015reducing}, where arterial theoretical waveforms are generated via a one-dimensional arterial network model. For the ease of comparison, all the waveforms have been normalized to a unitary amplitude and have been rescaled in time to last 0.8~s. 
  Fig.~\ref{fig:comparison_arterial_pressure} shows  {a good agreement between theoretical and experimental waveforms for the ascending aorta, the common carotid artery, the aortic arch and the thoracic aorta. The theoretical waveform predicted for the abdominal aorta agrees well with the results  by Davies et al, which appear to differ from Parker's results particularly in the systolic part of the cardiac cycle. The diastolic part of the theoretical waveform for the iliac artery agrees well with the one reported by Parker, whereas the theoretical systolic part appears to be shorter than the experimental ones. This difference might be due to the fact that our model lumps in a single compartment all the iliac segments, namely right and left common, external and internal iliac arteries.}

Overall, the results presented in this section provide evidence of the capability of the proposed closed-loop model to capture prominent features of cardiovascular physiology and support the utilization of the model output to construct a theoretical BCG, as discussed next. 

\subsection{Theoretical and experimental ballistocardiograms}
\label{sec:results_BCG}

The volume waveforms simulated using the closed-loop model (see Fig.~\ref{fig:model}) can be substituted in Eq.~(\ref{eq:BCG_disp})
to calculate theoretically the waveform BCG$_{disp}(t)$ associated with the displacement of the center of mass.
The BCG waveforms for velocity and acceleration, namely  BCG$_{vel}(t)$ and BCG$_{acc}(t)$,
are obtained from BCG$_{disp}(t)$
via discrete time differentiation.
%
Following~\cite{Noordergraaf1963,noordergraaf1958genesis,StarrNoordergraafBook},
the BCG waveforms are reported by means of the auxiliary functions $f_D$, $f_V$ and $f_A$ defined as 
\begin{align}
f_D(t) := & \,M \cdot \mbox{BCG}_{disp} (t) = \rho_b  \sum_{i=1}^N V_i(t) y_i & [\mbox{g cm}]\nonumber\\
f_V (t):= & \,M \cdot \mbox{BCG}_{vel} (t) = \rho_b  \sum_{i=1}^N \frac{dV_i(t)}{dt} y_i &[\mbox{g cm s}^{-1}]\nonumber\\
f_A(t) := & \,M \cdot \mbox{BCG}_{acc} (t) = \rho_b  \sum_{i=1}^N \frac{d^2V_i(t)}{dt^2}y_i &[\mbox{dyne}]\,.\nonumber
\end{align}
The auxiliary functions are independent of the value of the mass $M$ and, as a consequence, they allow for a fair comparison between BCG waveforms reported in different studies.
%
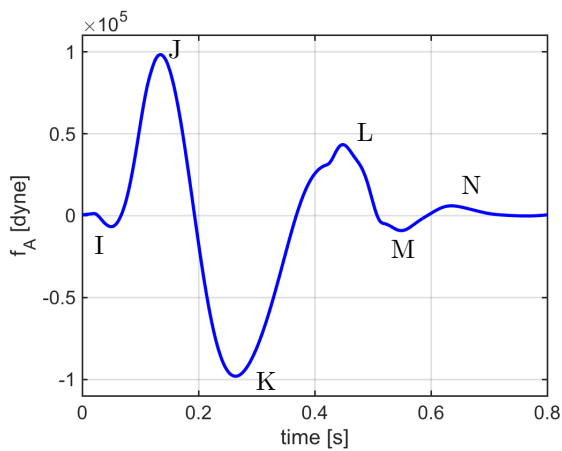
\begin{figure}[h!]
\centering
\begin{tikzpicture}[scale=1]
\node[inner sep=0pt] at (0,0)
    {\includegraphics[width=0.65\textwidth]{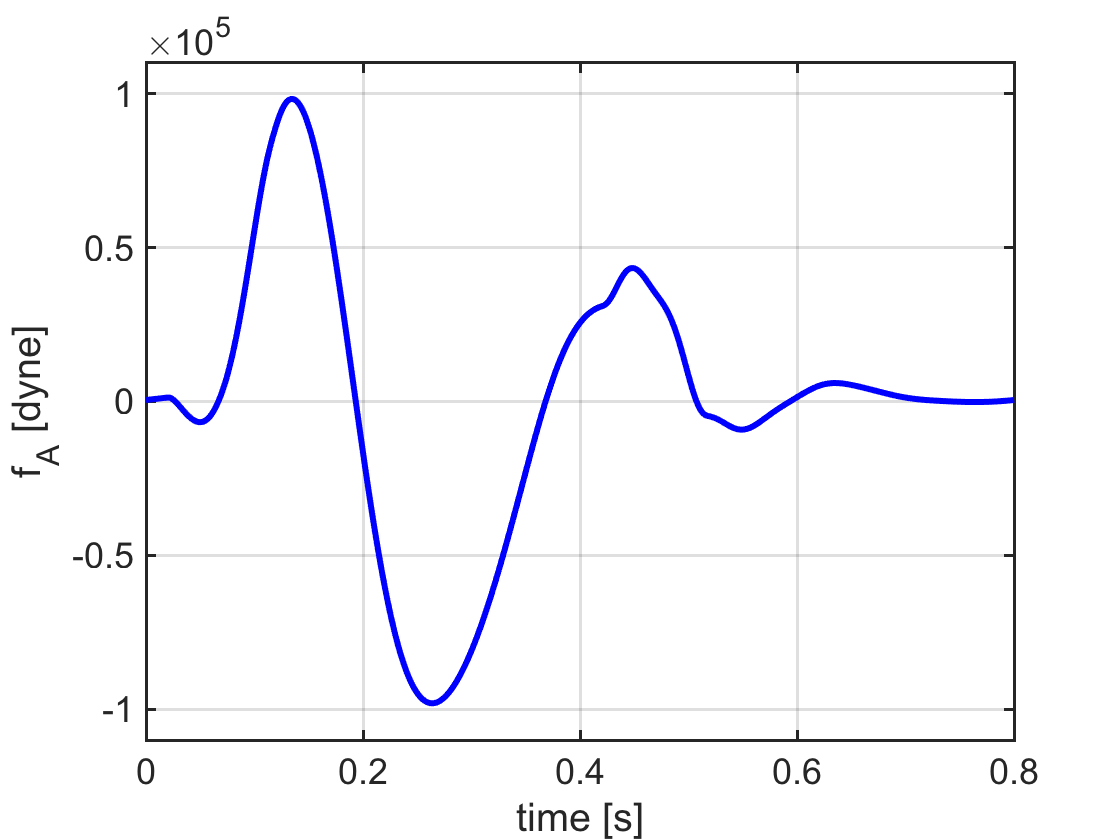}};
\node[above] at (-2.7,-0.5) {I};
\node[above] at (-1.7,2.1) {J};
\node[above] at (-0.5,-2.3) {K};
\node[above] at (0.8,1.) {L};
\node[above] at (1.3,-0.55) {M};
\node[above] at (2.2,0.3) {N};
\end{tikzpicture}
\caption{Waveform $f_A$ 
simulated via the closed-loop model over one cardiac cycle. The waveform exhibits the typical I, J, K, L, M and N peaks and valleys that characterize BCG signals measured experimentally.}
\label{fig:A_BCG_simulated}
\end{figure}

Fig.~\ref{fig:A_BCG_simulated} reports  $f_A(t)$ simulated via the closed-loop model over one cardiac cycle. The waveform exhibits the typical I, J, K, L, M and N peaks and valleys that characterize BCG signals measured experimentally~\cite{StarrNoordergraafBook,Kim2016,inan2015ballistocardiography}, thereby confirming the capability of the closed-loop model to capture the fundamental cardiovascular mechanisms that give rise to the BCG signal.

%
\begin{figure}[h!]
\centering
\includegraphics[width=0.85\textwidth]{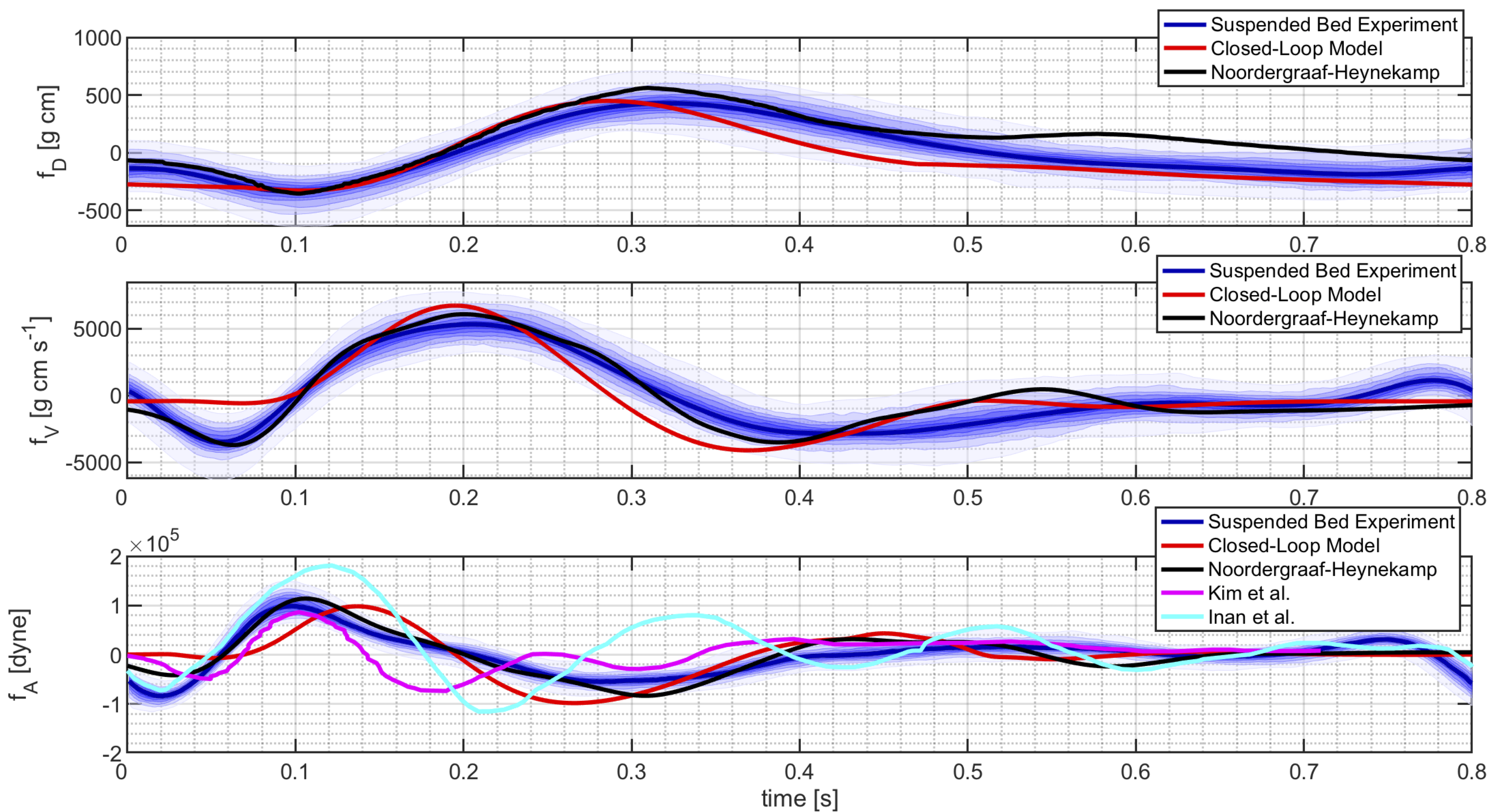}
\caption{BCG waveforms for displacement ($f_D$), velocity ($f_V$) and acceleration ($f_A$)  simulated via the closed-loop model are compared with the theoretical waveforms calculated by Noordergraaf and Heynekamp~\cite{noordergraaf1958genesis} and the experimental waveforms obtained by Inan et al~\cite{inan2015ballistocardiography},
Kim et al~\cite{Kim2016} and our group. 
}
\label{fig:Comparison_All_BCG}
\end{figure}


A quantitative comparison between the simulated BCG waveforms pertaining to displacement, velocity and acceleration of the center of mass is reported in Fig.~\ref{fig:Comparison_All_BCG} by means of the auxiliary functions $f_D$, $f_V$ and $f_A$. The waveforms simulated via the closed-loop model are compared with 
\emph{(i)} the theoretical results presented by Noordergraaf and Heynekamp~\cite{noordergraaf1958genesis}, where we calculated velocity and acceleration waveforms via time differentiation of the reported signal for the displacement;
\emph{(ii)} the experimental results obtained by Inan et al utilizing a modified weighing scale~\cite{inan2015ballistocardiography,inan2009robust};
\emph{(iii)} the experimental results obtained by Kim et al utilizing direct pressure measurements~\cite{Kim2016}; 
and 
\emph{(iv)} the experimental results obtained by our group utilizing an accelerometer on a suspended bed (see Section~\ref{sec:exp_data}), where we utilized the mass $M=87$Kg of the study subject to calculate $f_A$. A time shift has been applied in order to align the waveforms for ease of comparison.

The results  in Fig.~\ref{fig:Comparison_All_BCG} confirm that the BCG waveforms simulated via the closed-loop model are of the same order of magnitude as those reported in previous theoretical and experimental studies for all three aspects of the body motion, namely displacement, velocity and acceleration. The figure also shows that theoretical and experimental BCG waveforms obtained by different authors exhibit similar trends, typically marked by the peaks reported in Fig.~\ref{fig:A_BCG_simulated}, but differ in the precise timing of these peaks and in their magnitude. Such differences might be due to physiological variability of cardiovascular parameters among individuals, for example associated with age, gender and ethnicity~\cite{Maceira2006Left,Maceira2006Right,cattermole2017normal}, 
and to different techniques utilized to acquire the BCG signal, such as the modified weighing scale and the accelerometer on the suspended bed. 
In this perspective, the closed-loop model presented in this work could be used to perform a sensitivity analysis to identify the cardiovascular parameters that influence the BCG signal the most. In addition, the output of the proposed closed-loop model could be coupled with models describing the functioning principles of the specific measuring device in order to better interpret the acquired BCG signal.

\subsection{ {Towards a clinical interpretation of ballistocardiograms}}
\label{sec:pathological_conditions}
 {The proposed closed-loop model  can be used to predict how 
the BCG signal will be altered by
specific pathological conditions, thereby providing a quantitative method to identify and characterize BCG changes bearing specific clinical relevance. As an example, we simulate a reduced capability of the left ventricle to contract properly
by decreasing the value of the parameter $q_L$ in the activation function $a_L(t)$  of Eq. (\ref{eq:activ}). This condition is associated with \textit{left-sided heart failure with reduced ejection fraction} (HFrEF), also known as systolic failure. As another example, we simulate a reduced capability of the left ventricle to relax properly during diastole by increasing the value of the parameter ELD in the elastance function of Eq. (\ref{EL}). This condition is associated with \textit{left-sided heart failure with preserved ejection fraction} (HFpEF), also known as diastolic failure. We emphasize that we are not currently accounting for any remodeling phenomena that occur as heart failure persists; thus, the numerical simulations reported below are indicative of heart failure in its early stages.}

%
\begin{figure}[h!]
\centering
\includegraphics[width=0.85\textwidth]{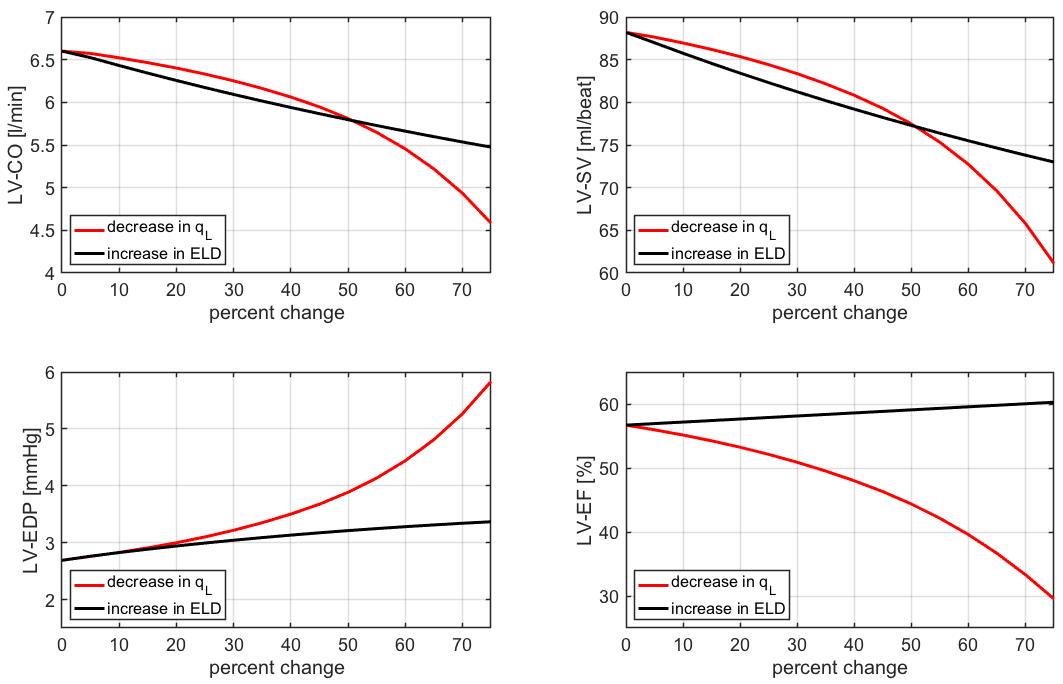}
\caption{ {Simulated values of cardiac output (CO), stroke volume (SV), end diastolic pressure (EDP)  and ejection fraction (EF) in the left ventricle (LV) for decrements in $q_L$ (\emph{red}) and increments in ELD (\emph{black}) up to 75\%, with baseline values corresponding to 0\% change.}
}
\label{fig:Cardiac_Failure_LV}
\end{figure}

\begin{figure}[h!]
\centering
\includegraphics[width=0.85\textwidth]{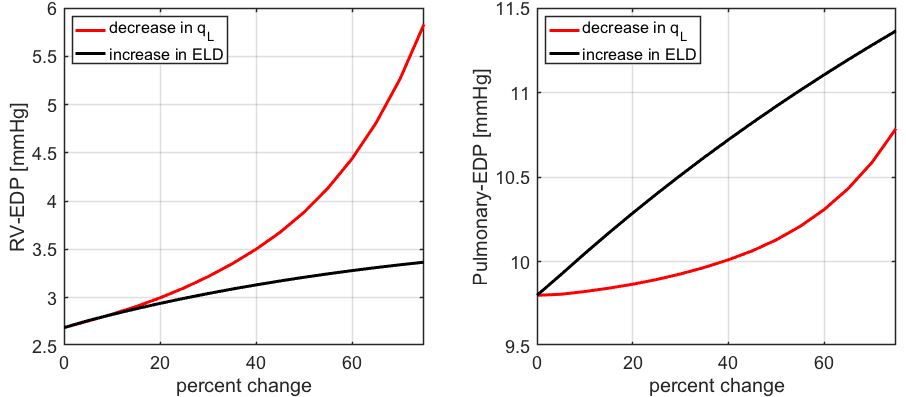}
\caption{ {Simulated values of end-diastolic pressure (EDP) in the right ventricle (RV) and the pulmonary artery for decrements in $q_L$ (\emph{red}) and increments in ELD (\emph{black}) up to 75\%, with baseline values corresponding to 0\% change.}
}
\label{fig:Cardiac_EDP}
\end{figure}

 {Fig.~\ref{fig:Cardiac_Failure_LV} reports the simulated values of cardiac output (CO), stroke volume (SV), end-diastolic pressure (EDP)  and ejection fraction (EF) in the left ventricle (LV) for decrements in $q_L$ (red curves) and increments in ELD (black curves) up to 75\% with respect to baseline. Here,   baseline values correspond to 0\% change. Regardless of whether $q_L$ is decreased or ELD is increased, the model predicts a decrease in CO and SV and an increase in EDP, as it happens in heart failure~\cite{schwartzenberg2012effects}. Furthermore, EF is predicted to decrease as $q_L$ decreases, as clinically observed in HFrEF, and to remain almost constant as ELD increases, as clinically observed in HFpEF~\cite{schwartzenberg2012effects}. }
 {As mentioned in Section~\ref{sec:background}, left-sided heart failure leads to an increase in fluid pressure in the lungs and  in the right ventricle, ultimately damaging the right side of the heart. Thanks to its closed-loop architecture, our cardiovascular model is able to capture this feedback phenomenon (Fig.~\ref{fig:Cardiac_EDP}), thereby holding promise to serve as a virtual laboratory where other complex pathologies of the cardiovascular system can be simulated and their manifestations on the BCG signal can be characterized.
}

%


%
\begin{figure}[h!]
\centering
\includegraphics[width=0.85\textwidth]{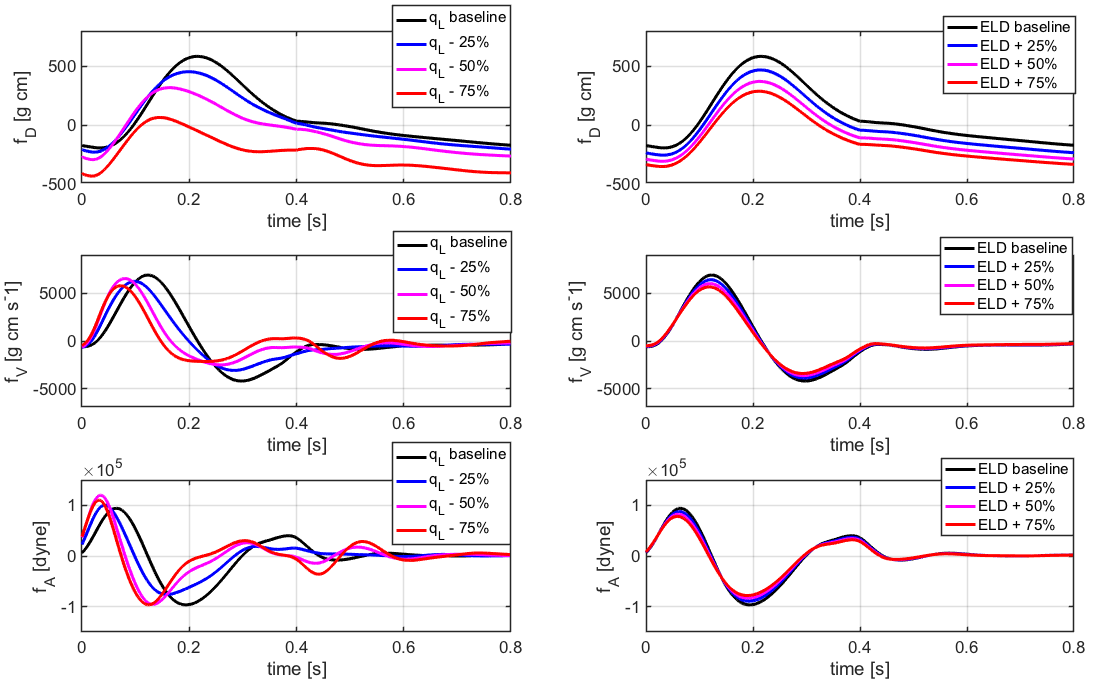}
\caption{ {Simulated BCG waveforms for displacement ($f_D$), velocity ($f_V$) and acceleration ($f_A$) associated with reduced $q_L$ and increased ELD are reported 
for baseline values (\emph{black}), 25\% changes (\emph{blue}), 50\% changes (\emph{magenta}) and 75\% changes (\emph{red}) in $q_L$ (\emph{left}) and ELD (\emph{right}). }
}
\label{fig:BCG_pathological}
\end{figure}


 {The simulated BCG waveforms associated with reduced $q_L$ and increased ELD are reported in Fig.~\ref{fig:BCG_pathological} by means of the auxiliary functions $f_D$, $f_V$ and $f_A$, which are indicative of displacement, velocity and acceleration of the center of mass of the human body. 
As $q_L$ decreases, the $f_D$ waveform exhibits major changes in shape including
 a flatter profile
and 
a time shift of its peaks and valleys with respect to baseline.
The shape changes in the $f_D$ waveform are naturally inherited by the $f_V$ and $f_A$ waveforms since they are obtained via time differentiation.
Conversely, as ELD increases, the $f_D$ waveform shows an overall decrease in magnitude without any major shape change. Thus, the $f_V$ and $f_A$ waveforms show only a slight change in the magnitude of peaks and valleys without major changes in their timing.}  {The information on the expected traits and magnitudes of BCG changes associated with specific  conditions can help optimize the design of BCG sensors in order to better capture changes in the signal that are indicative of specific pathologies. }
 {Several studies reported changes in BCG signals in relation to cardiac dysfunctions~\cite{giovangrandi2012preliminary,bruser2013automatic,etemadi2014tracking}. Our work provides a novel method to characterize BCG changes due to specific pathological conditions, thereby setting the foundations for a quantitative approach to the clinical interpretation of BCG signals.}

\section{Conclusions}
\label{sec:conclusions}

The proposed closed-loop model captures the predominant features of cardiovascular physiology giving rise to the BCG signal  {and provides a novel method to characterize BCG changes due to specific pathological conditions}. To the best of our knowledge, this work provides the first theoretical interpretation of BCG signals based on a
physically-based (biophysical), mathematical closed-loop model of the cardiovascular system. 
%
 {The proposed closed-loop model could be further improved by coupling it with existing}
open models for the arterial side of the circulation, e.g.~\cite{reymond2009validation,epstein2015reducing}.  {In addition, the model does not currently account for the natural variability of: 
\emph{(i)} relative motion of anatomical locations during the cardiac cycle; and
\emph{(ii)} model parameters among individuals. As a first step, we plan to perform a sensitivity analysis to identify the model parameters whose variation would induce significant changes in the BCG waveform, see e.g.~\cite{szopos2016mathematical}.}


The physically-based BCG modeling described in this paper integrates well with our prior work in monitoring older adults noninvasively and longitudinally in the home setting. Our in-home sensor system automatically generates health messages that indicate early signs of health change, thereby allowing for very early treatment, which has been shown to produce better health outcomes~\cite{rantz2015enhanced,rantz2015new}.
Several parameters are tracked currently, including in-home gait patterns~\cite{stone2013unobtrusive},
overall activity level~\cite{stone2013unobtrusive,stone2015average,skubic2015automated},
heart rate and respiration rate~\cite{rosales2017heart,lydon2015robust},
blood pressure~\cite{su2018monitoring},
and sleep patterns~\cite{Enayati2018,yang2016sleep}. 
 {
The proposed 
model will be used to enhance the 
data analysis systems for in-home sensors and improve the accuracy in the detection of deteriorating health conditions, in a continuing effort to provide better healthcare options for our aging population.}
%

\appendix
\section{Mathematical Formulation of Closed-Loop Model}\label{app:model}
The closed-loop model 
in Fig.~\ref{fig:model} can be written 
as a system of ordinary differential equations (ODEs) of the form
\begin{equation}\label{eq:system}
\mathcal M(Y(t)) \frac{d Y(t)}{dt} = \mathcal A (Y(t)) Y(t) + b(Y(t)) \quad t\in[0,T]
\end{equation}
where $Y$ is the $m-$dimensional column vector of unknowns, $\mathcal M$ and $\mathcal A$ are $m\times m$ tensors and $b$ is the $m-$dimensional column vector of given forcing terms. 
We anticipate that $\mathcal M$, $\mathcal A$ and $b$ nonlinearly 
depend on the vector of state variables $Y$.
The specific expressions for $Y$, $\mathcal M$, $\mathcal A$ and $b$ follow from the constitutive equations characterizing the circuit elements and the
 {Kirchhoff} laws of currents and voltages.

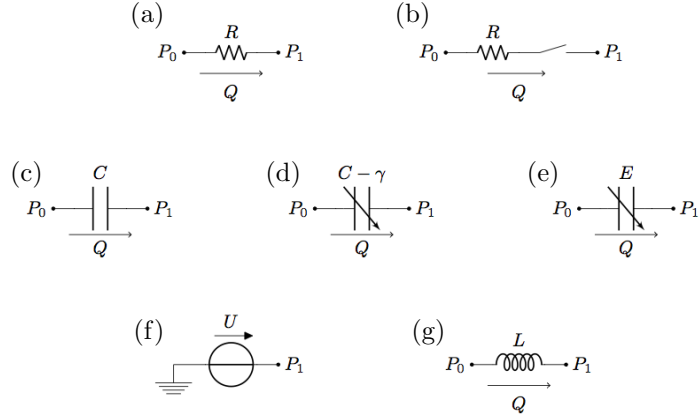
\begin{figure}[ht]
\centering
\begin{tikzpicture}
\node[inner sep=0pt] at (0,0)
    {\includegraphics[width=0.75\textwidth]{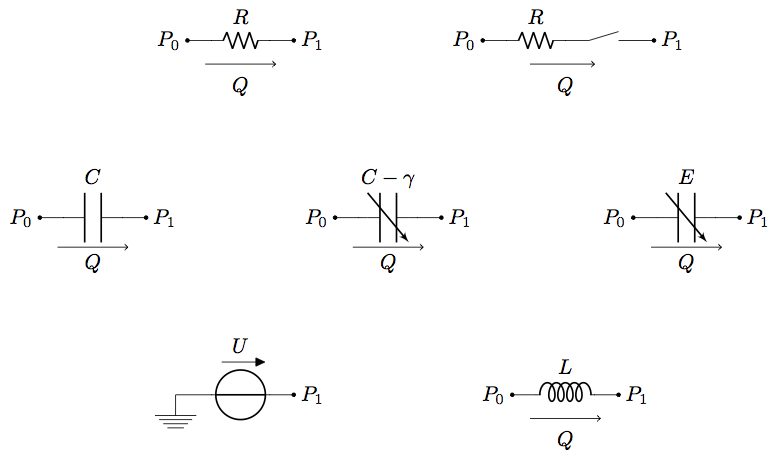}};
\node [left] at (-2.5,2.7) {(a)};
\node [left] at (1,2.7) {(b)};
\node [left] at (-4.1,0.6) {(c)};
\node [left] at (-0.7,0.6) {(d)};
\node [left] at (2.7,0.6) {(e)};
\node [left] at (-2.5,-1.5) {(f)};
\node [left] at (1.2,-1.5) {(g)};
\end{tikzpicture}
\caption{Electrical elements included in the closed-loop model: 
(a) linear resistor; 
(b) linear resistor with ideal switch; 
(c) linear capacitor; 
(d) variable capacitor for arterial viscoelasticity; 
(e) variable capacitor for ventricular elastance; 
(f) voltage source; 
(g) linear inductor.
}
\label{fig:elements}
\end{figure}


\begin{table*}[h!]
\caption{Parameter values for the closed-loop model}\label{tab:par_values}
\resizebox{\textwidth}{!}{
\begin{tabular}{lllll}
\hline\\
\textit{Heart}\\
 $T_c = 0.8$ s & $T_s = 0.4$ s & $T_a = 0.08$ s &$T_b = 0.45$ s \\
    {$q_L = 2\pi$} &  {$q_R = 2\pi$}&  
 ULO = 50 mmHg & ELD = 0.04 mmHg cm$^{-3}$ \\
 ELS = 1.375 mmHg cm$^{-3}$ s$^{-1}$ & $R_L= 0.008$ mmHg s  cm$^{-3}$&
 URO = 24 mmHg & ERD = 0.01 mmHg cm$^{-3}$ \\
  ERS = 0.23 mmHg cm$^{-3}$ s$^{-1}$ & $R_R= 0.0175$ mmHg s  cm$^{-3}$
\\[.07in]
\textit{Systemic Circulation}\\
$R_1=0.003751$ mmHg s  cm$^{-3}$ & $R_{2a} = 0.00003111$ mmHg s  cm$^{-3}$ & $R_{2b} = 0.00003111$ mmHg s  cm$^{-3}$ & $R_{3a} = 0.00011683$ mmHg s  cm$^{-3}$ \\
 $R_{3b} = 0.00011683$ mmHg s  cm$^{-3}$ &
  $R_{4a} = 0.00061427$ mmHg s  cm$^{-3}$ &  $R_{4b} = 0.00061427$ mmHg s  cm$^{-3}$ &
    $R_{5a} = 0.00101868$ mmHg s  cm$^{-3}$ \\
      $R_{5b} = 0.00101868$ mmHg s  cm$^{-3}$ &
      $R_{6a} = 0.00265297$ mmHg s  cm$^{-3}$ &  $R_{6b} = 0.00265297$ mmHg s  cm$^{-3}$ &
      $R_{7} = 0.35$ mmHg s  cm$^{-3}$ \\
      $R_{8} = 0.0675$ mmHg s  cm$^{-3}$ &
      $R_{9} = 2.0$ mmHg s  cm$^{-3}$ &  $R_{10} = 0.003751$ mmHg s  cm$^{-3}$ &
       \\
       $C_2 = 0.13853688$ cm$^{3}$ mmHg$^{-1}$ & $\gamma_2 = 0.00713074$ mmHg s cm$^{-3}$& $C_3 = 0.12078980$ cm$^{3}$ mmHg$^{-1}$& $\gamma_3 = 0.08117842$ mmHg s cm$^{-3}$\\
              $C_4 = 0.21968142$ cm$^{3}$ mmHg$^{-1}$ & $\gamma_4 = 0.00449683$ mmHg s cm$^{-3}$& $C_5 = 0.17355893$ cm$^{3}$ mmHg$^{-1}$& $\gamma_5 = 0.00569184$ mmHg s cm$^{-3}$\\
              $C_6 = 0.02062154$ cm$^{3}$ mmHg$^{-1}$ & $\gamma_6 = 0.04790476$ mmHg s cm$^{-3}$ & $C_7 = 0.8$ cm$^{3}$ mmHg$^{-1}$  & $C_8 = 1.46 $ cm$^{3}$ mmHg$^{-1}$ \\
              $C_9 = 20.0 $ cm$^{3}$ mmHg$^{-1}$ \\
              $L_3 = 0.00113873$ mmHg s$^{2}$  cm$^{-3}$ & $L_4 = 0.00424581 $ mmHg s$^{2}$  cm$^{-3}$& $L_5 = 0.00551999$ mmHg s$^{2}$  cm$^{-3}$& $L_6 = 0.00538022$mmHg s$^{2}$  cm$^{-3}$ \\
              $L_7 = 0.000225$ mmHg s$^{2}$  cm$^{-3}$& $L_8 = 0.000225$ mmHg s$^{2}$  cm$^{-3}$& $L_9 = 0.0036$  mmHg s$^{2}$  cm$^{-3}$\\[.07in]
\textit{Pulmonary Circulation}\\
$R_{11}=0.003751$ mmHg s  cm$^{-3}$ & $R_{12} = 0.03376 $ mmHg s  cm$^{-3}$ & $R_{13a} = 0.1013$ mmHg s  cm$^{-3}$ & $R_{13b} = 0.003751$ mmHg s  cm$^{-3}$ \\
 $C_{11} = 0.09 $ cm$^{3}$ mmHg$^{-1}$ &  $C_{12} = 2.67 $ cm$^{3}$ mmHg$^{-1}$ &  $C_{13} = 46.7 $ cm$^{3}$ mmHg$^{-1}$ \\
 $L_{12} = 0.00075 $  mmHg s$^{2}$  cm$^{-3}$ & $L_{13} = 0.00308$  mmHg s$^{2}$  cm$^{-3}$ \\[.07in]
 \textit{Cerebral Circulation}\\
$R_{14a}=0.03006924 $ mmHg s  cm$^{-3}$ & $R_{14b} = 0.03006924  $ mmHg s  cm$^{-3}$ & $R_{cap1} = 0.327$ mmHg s  cm$^{-3}$ & $R_{cap2} = 0.327$ mmHg s  cm$^{-3}$ \\
$R_{15a}=0.327$ mmHg s  cm$^{-3}$ & $R_{15b} = 0.327 $ mmHg s  cm$^{-3}$  \\
 $C_{14} = 0.03790125 $ cm$^{3}$ mmHg$^{-1}$ &  $\gamma_{14} = 0.04110141$ mmHg s cm$^{-3}$ &  $C_{15} = 0.688 $ cm$^{3}$ mmHg$^{-1}$ \\
 $L_{14} = 0.03430149 $  mmHg s$^{2}$  cm$^{-3}$ & $L_{cap} = 0.00424581$  mmHg s$^{2}$  cm$^{-3}$ 
 & $L_{15} = 0.00424581$  mmHg s$^{2}$  cm$^{-3}$ \\[.07in]
 \hline
\end{tabular}}
\end{table*}

The types of electrical elements included in the closed-loop model are summarized in Fig.~\ref{fig:elements}. The constitutive laws characterizing each element are detailed below.
\begin{subequations}
\label{eq:constitutive_laws_electric_elements} 
\begin{itemize}
\item \emph{Linear resistor} (Fig.~\ref{fig:elements}(a)): the hydraulic analog to Ohm's law states that the volumetric flow rate $Q$
is proportional to the pressure difference $P_0-P_1$ across the resistor, namely 
\begin{equation}
Q=\frac{(P_0-P_1)}{R},
\end{equation}
where $R$ is the hydraulic resistance.
\item \emph{Linear resistor with ideal switch} (Fig.~\ref{fig:elements}(b)): in this element, often  utilized  to model heart valves~\cite{Avanzolini1988,formaggia2010cardiovascular},  the switch is  completely open as soon as the pressure difference $P_0-P_1$ is positive and completely closed otherwise, namely
\begin{equation}
Q = \sigma_{P_0-P_1} \frac{(P_0-P_1)}{R},
\end{equation}
where 
$\sigma_{P_0-P_1} = 1$ if $P_0-P_1>0$, equal to 0 otherwise.
%
\item \emph{Linear capacitor} (Fig.~\ref{fig:elements}(c)): the time rate of change of the fluid volume $V$ stored in a capacitor equals the volumetric flow rate $Q$, namely
\begin{equation}\label{eq:capacitor}
\frac{d V}{dt} = Q\,.
\end{equation}
In the case of linear capacitor, the volume $V$ and the
pressure difference $P_0-P_1$ are related by a proportionality law, namely 
$V = C (P_0-P_1)$,
where $C$ is a positive constant called capacitance.
\item \emph{Variable capacitor for arterial viscoelasticity} (Fig.~\ref{fig:elements}(d)): the law for the capacitor stated by
Eq.~(\ref{eq:capacitor}) is coupled with the following differential relationship between 
the pressure difference $P_0-P_1$ and the fluid volume $V$
\begin{equation}\label{eq:vC}
P_0-P_1 = \frac{V}{C} + \gamma \frac{dV}{dt}
\end{equation}
where $C$ and $\gamma$ are positive constants. Relationship~(\ref{eq:vC}) corresponds to adopting a linear viscoelastic thin shell model for the arterial walls, as in~\cite{Canic2006,vcanic2006blood}. 
\item \emph{Variable capacitor for ventricular elastance} (Fig.~\ref{fig:elements}(e)): 
the law for the capacitor stated by
Eq.~(\ref{eq:capacitor}) is coupled with a relationship between
the pressure difference $P_0-P_1$ and the fluid volume $V$ of the form
$
P_0-P_1 = E(t)\, V
$,
where $E(t)$ is a given function of time modeling the complex biomechanical properties of the ventricular wall. Following~\cite{Avanzolini1988}, we assume
\begin{eqnarray}
E_L(t) &= \mbox{ELD}+\mbox{ELS}\,a
_L(t), \label{EL}\\
 E_R(t) &= \mbox{ERD}+\mbox{ERS}\,a_R(t),
\end{eqnarray}
where ELD, ELS, ERD and ERS are given constants  {characterizing the systolic and diastolic elastances of the left and right ventricles, whereas $a_L(t)$ and $a_R(t)$ are the activation functions characterizing the ventricle contractions~\cite{Avanzolini1988}. The activation functions are defined as
\begin{equation}\label{eq:activ}
\begin{array}{c}
a_L(t) =\displaystyle\frac{\tanh(q_L \, t_a) - \tanh(q_L \,t_b )}{2}\\[.1in]
a_R(t) =\displaystyle\frac{\tanh(q_R \, t_a) - \tanh(q_R \,t_b )}{2}
\end{array}
\end{equation}
if $t_m<T_s$, where $T_s$ is the length of the systolic part of the cardiac cycle,
  and equal to 0 otherwise. Here, $t_m = \mod(t,T_c)$, $t_a = t_m-T_a$ and $t_b = t_m-T_b$,
where $T_c$ is the length of one cardiac cycle and $T_a$ and $T_b$ are given time constants characterized via electrocardiography. Specifically, $T_a$ corresponds to the T wave peak time and $T_b$ corresponds to the T wave offset time~\cite{le2013real} with respect to the R wave peak. }
\item \emph{Voltage source} (Fig.~\ref{fig:elements}(f)): the hydraulic analog of a voltage source is an element that imposes the nodal pressure as
$P_1(t) = U(t)$,
where $U(t)$ is a given function. Following~\cite{Avanzolini1988}, we assume
$
U_L(t) = \mbox{ULO}\, {a_L(t)}
$ and
$U_R(t) = \mbox{URO}\, {a_R(t)}$,
where ULO and URO are positive constants and   {$a_L(t)$ and $a_R(t)$} are given in Eq. (\ref{eq:activ}).
\item \emph{Linear inductor} (Fig.~\ref{fig:elements}(f)): the time rate of change of the volumetric flow rate $Q$ is related to the pressure difference $P_0-P_1$ by the following proportionality law 
\begin{equation}
L\frac{dQ}{dt} = P_0-P_1
\end{equation}
where $L$ is a positive constant.
\end{itemize}
\end{subequations}

The parameter values for each of the circuit elements are reported in Table~\ref{tab:par_values}. 
The parameter values pertaining to the heart, the systemic microcirculation and the pulmonary circulation have been adapted from~\cite{Avanzolini1988}. The parameter values pertaining to the main arteries have been computed using the following constitutive equations: 
\begin{equation*}
\label{eq:RCL_parameters}
	R = \dfrac{8\, \pi \, l \, \eta}{S^2}, \quad
	L = \dfrac{\rho_b \, l}{S}, \quad
	C = \dfrac{3 \, l\, S \, (a+1)^2}{E \, (2a+1)}, \quad
	\gamma = \dfrac{\delta}{C}, 
\end{equation*}
where $a=r/h$ is the ratio between vessel radius $r$ and wall thickness $h$, $l$ is the vessel length, $S=\pi r^2$ is the vessel cross-sectional area, $\rho_b$ is the blood density,  $\eta$ is the blood viscosity, $E$ and $\delta$ are the Young modulus and the viscoelastic parameter characterizing the vessel wall. 
In this work, we have assumed
$\rho_b = 1.05$ g cm$^{-3}$,
$\eta = 0.035$ g cm$^{-1}$ s$^{-1}$,
$E = 4 \cdot10^6$ dyne cm$^{-2}$ and
$\delta = 1.56 \cdot10^{-3}$ s.   
The values of the remaining geometrical  parameters utilized to determine $R$, $L$, $C$ and $\gamma$ for each of the main arterial segments 
have been adapted from \cite{Noordergraaf1963} and \cite{Canic2006} and
are reported in Table~\ref{tab:RCL_parameters}.

\begin{table}[h!]
\caption{Geometrical parameters for the main arterial segments}
\label{tab:RCL_parameters}
\centering
\begin{tabular}{llll}
\hline\\
\textsc{Arterial Segment} & $l$ [cm] & $r$ [cm] & $h$ [cm] \\
\hline\\
Ascending Aorta 	& 4  		& 1.44 & 0.158\\
Aortic Arch		& 5.9		& 1.25 & 0.139\\
Thoracic Aorta 		& 15.6	& 0.96 & 0.117\\
Abdominal Aorta 	& 15.9	& 0.85 & 0.105\\
Iliac artery 		& 5.8		& 0.52 & 0.076\\
Carotid artery 		& 20.8	& 0.39 & 0.064\\
\hline
\end{tabular}
\end{table}

Let us define the vector $Y(t)$ of the circuit unknowns in Eq.~(\ref{eq:system}) as 
the column vector 
\begin{equation}
Y(t) = [\mathcal V(t);\mathcal Q(t)]^T,
\end{equation}
where the two row vectors $\mathcal V$ and $\mathcal Q$ are defined as
\begin{align}
\mathcal V(t) =& [\,V_L(t), V_2(t), V_3(t), V_4(t), V_5(t), V_6(t), V_7(t), V_8(t), \nonumber\\[.05in]
&\,\; V_9(t), V_R(t), V_{11}(t), V_{12}(t), V_{13}(t), V_{14}(t), V_{15}(t)\,] \nonumber \\[.1in]
\mathcal Q(t) =& [\,Q_3(t), Q_4(t), Q_5(t), Q_6(t), Q_7(t), Q_8(t), \nonumber\\[.05in]
&\,\; Q_9(t), Q_{12}(t), Q_{13}(t), Q_{14}(t), Q_{15}(t)\,]. \nonumber
\end{align}
The symbols $V_L$ and $V_R$ denote the fluid volume stored in the variable capacitors for ventricular elastance
characterized by $E_L$ and $E_R$, respectively; 
the symbols $V_i$, $i=2, \ldots, 6$ and $i=14$
 denote the fluid volume stored in the variable capacitors for arterial viscoelasticity characterized by
 $C_i$ and $\gamma_i$;
 the symbols $V_i$, $i=7, \ldots, 9$ and $i=15$
  denote the fluid volume stored in the linear capacitors
characterized by $C_i$; the symbols 
$Q_i$, $i=3, \ldots, 9$ and $i= 12, \ldots, 15$, denote the volumetric flow rate 
through the inductor characterized by $L_i$.
The nonlinear system of ODEs~\eqref{eq:system} is derived by:
\begin{enumerate}
\item writing  {Kirchhoff's} current laws (KCLs)  for each of the 15 nodes marked on the circuit in Fig.~\ref{fig:model}; 
\item writing  {Kirchhoff's} voltage laws (KVLs)  for each of the 11 circuit branches containing an inductor;
\item substituting the constitutive equations~\eqref{eq:constitutive_laws_electric_elements} 
in the KCLs and KVLs.
\end{enumerate}

\begin{table}[h!]
\caption{Initial conditions for the time-dependent model simulations}
\label{tab:init_cond}
\centering
\begin{tabular}{ll|ll}
\hline
\textsc{Variable} & \textsc{Value} & \textsc{Variable} & \textsc{Value} \\
\hline 
$V_L$ &  71.2700 cm$^{3}$ &  $Q_3$ &  1.68 cm$^{3}$ s$^{-1}$  \\
$V_2$ &  73.5486 cm$^{3}$ & $Q_4$ &  1.9961 cm$^{3}$ s$^{-1}$ \\
$V_3$ &  71.9746 cm$^{3}$ & $Q_5$ &  1.1861 cm$^{3}$ s$^{-1}$ \\
$V_4$ &  71.9983 cm$^{3}$ &$Q_6$ &  9.03697 cm$^{3}$ s$^{-1}$  \\
$V_5$ &  71.9327 cm$^{3}$ & $Q_7$ &  17.8121 cm$^{3}$ s$^{-1}$\\
$V_6$ &  72.2213 cm$^{3}$ &  $Q_8$ &  19.1462 cm$^{3}$ s$^{-1}$ \\
$V_7$ &  80.9077 cm$^{3}$ & $Q_9$ &  67.359 cm$^{3}$ s$^{-1}$\\
$V_8$ &  70.537 cm$^{3}$ & $Q_{12}$ &  0.7861 cm$^{3}$ s$^{-1}$ \\
$V_9$ &  3.3268 cm$^{3}$ & $Q_{13}$ &  23.83 cm$^{3}$ s$^{-1}$ \\
$V_R$ &  3.1638 cm$^{3}$ & $Q_{14}$ &  0.5909 cm$^{3}$ s$^{-1}$  \\
$V_{11}$ &  13.4160 cm$^{3}$ &$Q_{15}$ &  1.9961 cm$^{3}$ s$^{-1}$   \\
$V_{12}$ &  13.3920 cm$^{3}$ & \\
$V_{13}$ &  11.2950cm$^{3}$&  \\
$V_{14}$ &  70.9869  cm$^{3}$& \\
$V_{15}$ &  3.3268 cm$^{3}$\\
\hline
\end{tabular}
\end{table}

The system is solved using the initial conditions reported in Table~\ref{tab:init_cond} and simulations are run until a periodic solution is established. 
Overall, the differential system~\eqref{eq:system} includes $m=26$ differential equations.
The expressions of the nonzero entries of the matrices $\mathcal M$ and $\mathcal A$
as well as of the forcing vector $b$ are reported below.
Let $\widetilde{R}_L:= R_L+R_1+R_{2a}$, $\widetilde{R}_R:= R_R+R_{11}$, 
$\widetilde{R}_{cap}:= R_{14b}+ R_{cap1}+ R_{cap2} +R_{15a}$ and $\widetilde{L}_{cap}:= L_{cap}+L_{15}$.
The nonzero entries of $\mathcal M$ are:
\begin{align*}
& \mathcal M_{1,1} = 1, \quad \mathcal M_{12} = -\Frac{\gamma_2}{\widetilde{R}_L}\sigma_{P_1-P_2} & \\
& \mathcal M_{2,2} = 1 + \Frac{\gamma_2}{\widetilde{R}_L}\sigma_{P_1-P_2} & \\
& \mathcal M_{i,i} = 1 \quad i=3, \ldots, 15 & \\
& \mathcal M_{i,j} = \gamma_j, \, \mathcal M_{i,j+1} = -\gamma_{j+1}, \,
\mathcal M_{i,i} = -L_{j+1}, & \nonumber \\
& \, i=16, \ldots, 19, \, j=i-14, &\\
& \mathcal M_{20,6} = \gamma_6, \quad \mathcal M_{20,20} = -L_7 & \\
& \mathcal M_{i,i} = -L_k, \quad i=21, 22, \quad k=i-13, & \\
& \mathcal M_{i,i} = -L_k, \quad i=23, 24 , \quad k=i-11, & \\
& \mathcal M_{25,3} = \gamma_3, \quad \mathcal M_{25,14} = -\gamma_{14}, \quad
\mathcal M_{25,25} = -L_{14} & \\
& \mathcal M_{26,14} = \gamma_{14}, \quad \mathcal M_{26,26} = -\widetilde{L}_{cap}. & 
\end{align*}
The nonzero entries of $\mathcal A$ are:
\begin{align*}
& \mathcal A_{1,1} = -E_L \left( \Frac{\sigma_{P_{13}-P_1}}{R_{13b}} + \Frac{\sigma_{P_{1}-P_2}}
{\widetilde{R}_{L}}\right) & \\
& \mathcal A_{1,2} = \Frac{\sigma_{P_{1}-P_2}}{\widetilde{R}_{L} C_{2}}, \quad
\mathcal A_{1,13} = \Frac{\sigma_{P_{13}-P_1}}{R_{13b} C_{13}} & \\
& \mathcal A_{2,1} = E_L \Frac{\sigma_{P_{1}-P_2}}{\widetilde{R}_{L}}, \quad
\mathcal A_{2,2} = -\Frac{\sigma_{P_{1}-P_2}}{\widetilde{R}_{L} C_{2}}, \quad
\mathcal A_{2,16} = -1 & \\
& \mathcal A_{3,16} = 1, \quad \mathcal A_{3,17}=-1, \quad \mathcal A_{3,25}=-1 & \\
& \mathcal A_{i,j} = 1, \quad  \mathcal A_{i,j+1} = -1 \qquad i=4, \ldots, 8, \, j=i+13  &\\
& \mathcal A_{9,9} = -\Frac{1}{C_9} \left( \Frac{\sigma_{P_{9}-P_{10}}}{R_{10}} + 
\Frac{1}{R_{15b}}\right) 
& \\
& \mathcal A_{9,10} = \Frac{E_R}{R_{10}}\sigma_{P_{9}-P_{10}}, \quad 
\mathcal A_{9,15} = \Frac{1}{R_{15b} C_{15}}, 
\quad 
\mathcal A_{9, 22} =1 & \\
& \mathcal A_{10,9} = \Frac{\sigma_{P_{9}-P_{10}}}{R_{10} C_9}, \quad
\mathcal A_{10,10} = - E_R \left( \Frac{\sigma_{P_{9}-P_{10}}}{R_{10}} + 
\Frac{\sigma_{P_{10}-P_{11}}}{\widetilde{R}_R}\right)& \\
& \mathcal A_{10,11} = \Frac{\sigma_{P_{10}-P_{11}}}{\widetilde{R}_R C_{11}},\quad 
\mathcal A_{11,10} = E_R \Frac{\sigma_{P_{10}-P_{11}}}{\widetilde{R}_R}
& \\
&
\mathcal A_{11,11} = - \Frac{\sigma_{P_{10}-P_{11}}}{\widetilde{R}_R C_{11}}, \quad 
\mathcal A_{11,23} = -1,\quad
\mathcal A_{12,23}=1  & \\
& 
\mathcal A_{12,24} = -1, \quad 
\mathcal A_{13,1} = E_L\Frac{\sigma_{P_{13}-P_{1}}}{R_{13b}}, \quad
\mathcal A_{13,13} = -\Frac{\sigma_{P_{13}-P_{1}}}{R_{13b} C_{13}}
& \\
& 
\mathcal A_{13,24} = 1, \quad 
\mathcal A_{14,25} = 1, \qquad  \mathcal A_{14,26} = -1 & \\
& \mathcal A_{15,9} = \Frac{1}{R_{15b} C_9}, \, \, \mathcal A_{15,15} = - \Frac{1}{R_{15b} C_{15}}, 
\, \, \mathcal A_{15,26} = 1 & \\
& \mathcal A_{i,j} = -\Frac{1}{C_j}, \quad \mathcal A_{i,j+1} = \Frac{1}{C_{j+1}}, \quad
\mathcal A_{i, i} = (R_{jb} + R_{j+1a}), & \nonumber \\
&  i=16, \ldots, 19, \, j=i-14  
&\\
& \mathcal A_{20,6} = -\Frac{1}{C_6}, \quad \mathcal A_{20,7} = \Frac{1}{C_7}, \quad
\mathcal A_{20,20} = (R_7 + R_{6b}) & \\
& \mathcal A_{i,j} = -\Frac{1}{C_j}, \quad \mathcal A_{i,j+1} = \Frac{1}{C_{j+1}}, \quad
\mathcal A_{i, i} = R_{j+1}, 
& \nonumber \\
&  i=21, 22, \, j=i-14, 
&\\
& \mathcal A_{23,11} = - \Frac{1}{C_{11}}, \quad \mathcal A_{23,12} = \Frac{1}{C_{12}},
\quad \mathcal A_{23,23} = R_{12} & \\
& \mathcal A_{24,12} = - \Frac{1}{C_{12}}, \quad \mathcal A_{24,13} = \Frac{1}{C_{13}},
\quad \mathcal A_{24,24} = R_{13a} & \\
& \mathcal A_{25,3} = - \Frac{1}{C_{3}}, \quad \mathcal A_{25,14} = \Frac{1}{C_{14}},
\quad \mathcal A_{25,25} = R_{14a} & \\
& \mathcal A_{26,14} = - \Frac{1}{C_{14}}, \quad \mathcal A_{26,15} = \Frac{1}{C_{15}},
\quad \mathcal A_{26,26} = \widetilde{R}_{cap}. & 
\end{align*}
The nonzero entries of the forcing term $b$ are:
\begin{align*}
& b_1 = -U_L \left( \Frac{\sigma_{P_{13}-P_1}}{R_{13b}} + \Frac{\sigma_{P_{1}-P_2}}
{\widetilde{R}_{L}}\right) & \\
& b_2 =  \Frac{U_L}{\widetilde{R}_{L}} \sigma_{P_{1}-P_2}, 
\quad b_9 =  \Frac{U_R}{R_{10}} \sigma_{P_{9}-P_{10}} 
& \\
& b_{10} = -U_R \left( \Frac{\sigma_{P_{9}-P_{10}}}{R_{10}} + \Frac{\sigma_{P_{10}-P_{11}}}
{\widetilde{R}_{R}}\right) & \\
& b_{11} = \Frac{U_R}{\widetilde{R}_{R}} \sigma_{P_{10}-P_{11}}, \quad 
b_{13} = \Frac{U_L}{R_{13b}} \sigma_{P_{13}-P_{1}}. 
& \\
\end{align*}

\section*{Acknowledgment}

The authors acknowledge funds from the University of Missouri and the Center for Eldercare and Rehabilitation Technology.


\bibliographystyle{plain}
\bibliography{../IEEEtran/bcg_biblio.bib}

\begin{thebibliography}{10}
\providecommand{\url}[1]{#1}
\csname url@samestyle\endcsname
\providecommand{\newblock}{\relax}
\providecommand{\bibinfo}[2]{#2}
\providecommand{\BIBentrySTDinterwordspacing}{\spaceskip=0pt\relax}
\providecommand{\BIBentryALTinterwordstretchfactor}{4}
\providecommand{\BIBentryALTinterwordspacing}{\spaceskip=\fontdimen2\font plus
\BIBentryALTinterwordstretchfactor\fontdimen3\font minus
  \fontdimen4\font\relax}
\providecommand{\BIBforeignlanguage}[2]{{%
\expandafter\ifx\csname l@#1\endcsname\relax
\typeout{** WARNING: IEEEtran.bst: No hyphenation pattern has been}%
\typeout{** loaded for the language `#1'. Using the pattern for}%
\typeout{** the default language instead.}%
\else
\language=\csname l@#1\endcsname
\fi
#2}}
\providecommand{\BIBdecl}{\relax}
\BIBdecl

\bibitem{StarrNoordergraafBook}
I.~Starr and A.~Noordergraaf, \emph{Ballistocardiography in cardiovascular
  research: Physical aspects of the circulation in health and disease}.\hskip
  1em plus 0.5em minus 0.4em\relax Lippincott, 1967.

\bibitem{Pinheiro2010}
E.~Pinheiro~et al, ``Theory and developments in an unobtrusive cardiovascular
  system representation: ballistocardiography.''

\bibitem{starr1961twenty}
I.~Starr and F.~C. Wood, ``Twenty-year studies with the ballistocardiograph:
  the relation between the amplitude of the first record of healthy adults and
  eventual mortality and morbidity from heart disease,'' \emph{Circulation},
  vol.~23, no.~5, pp. 714--732, 1961.

\bibitem{inan2015ballistocardiography}
O.~T. Inan, P.-F. Migeotte, K.-S. Park, M.~Etemadi, K.~Tavakolian,
  R.~Casanella, J.~M. Zanetti, J.~Tank, I.~Funtova, G.~K. Prisk \emph{et~al.},
  ``Ballistocardiography and seismocardiography: a review of recent advances.''
  \emph{IEEE Journal of Biomedical and Health Informatics}, vol.~19, no.~4, pp.
  1414--1427, 2015.

\bibitem{shin2008automatic}
J.~Shin, B.~Choi, Y.~Lim, D.~Jeong, and K.~Park, ``Automatic ballistocardiogram
  ({B}{C}{G}) beat detection using a template matching approach,'' in
  \emph{Engineering in Medicine and Biology Society, 2008. EMBS 2008. 30th
  Annual International Conference of the IEEE}.\hskip 1em plus 0.5em minus
  0.4em\relax IEEE, 2008, pp. 1144--1146.

\bibitem{inan2009robust}
O.~Inan, M.~Etemadi, R.~Wiard, L.~Giovangrandi, and G.~Kovacs, ``Robust
  ballistocardiogram acquisition for home monitoring,'' \emph{Physiological
  Measurement}, vol.~30, no.~2, p. 169, 2009.

\bibitem{giovangrandi2011ballistocardiography}
L.~Giovangrandi, O.~T. Inan, R.~M. Wiard, M.~Etemadi, and G.~T. Kovacs,
  ``Ballistocardiography - a method worth revisiting,'' in \emph{Engineering in
  Medicine and Biology Society, EMBC, 2011 Annual International Conference of
  the IEEE}.\hskip 1em plus 0.5em minus 0.4em\relax IEEE, 2011, pp. 4279--4282.

\bibitem{rate2007emfit}
R.~Rate, ``Emfit sensor technology,'' \emph{Journal of Medical Systems},
  vol.~31, pp. 69--77, 2007.

\bibitem{alametsa2008ballistocardiography}
J.~Alamets{\"a}, J.~Viik, J.~Alakare, A.~V{\"a}rri, and A.~Palom{\"a}ki,
  ``Ballistocardiography in sitting and horizontal positions,''
  \emph{Physiological Measurement}, vol.~29, no.~9, p. 1071, 2008.

\bibitem{paalasmaa2015adaptive}
J.~Paalasmaa, H.~Toivonen, M.~Partinen \emph{et~al.}, ``Adaptive heartbeat
  modeling for beat-to-beat heart rate measurement in ballistocardiograms.''
  \emph{IEEE Journal of Biomedical and Health Informatics}, vol.~19, no.~6, pp.
  1945--1952, 2015.

\bibitem{chen2008unconstrained}
W.~Chen, X.~Zhu, T.~Nemoto, K.-i. Kitamura, K.~Sugitani, and D.~Wei,
  ``Unconstrained monitoring of long-term heart and breath rates during
  sleep,'' \emph{Physiological Measurement}, vol.~29, no.~2, p.~N1, 2008.

\bibitem{katz2016contact}
Y.~Katz, R.~Karasik, and Z.~Shinar, ``Contact-free piezo electric sensor used
  for real-time analysis of inter beat interval series,'' in \emph{Computing in
  Cardiology Conference (CinC), 2016}.\hskip 1em plus 0.5em minus 0.4em\relax
  IEEE, 2016, pp. 769--772.

\bibitem{rosales2017heart}
L.~Rosales, B.~Y. Su, M.~Skubic, and K.~Ho, ``Heart rate monitoring using
  hydraulic bed sensor ballistocardiogram,'' \emph{Journal of Ambient
  Intelligence and Smart Environments}, vol.~9, no.~2, pp. 193--207, 2017.

\bibitem{huffaker2018passive}
M.~F. Huffaker, M.~Carchia, B.~U. Harris, W.~C. Kethman, T.~E. Murphy, C.~C.
  Sakarovitch, F.~Qin, and D.~N. Cornfield, ``Passive nocturnal physiologic
  monitoring enables early detection of exacerbations in asthmatic children: A
  proof of concept study,'' \emph{American Journal of Respiratory and Critical
  Care Medicine}, no.~ja, 2018.

\bibitem{young2008method}
S.~J. Young, W.~M. Gillon, R.~V. Rifredi, and W.~T. Krein, ``Method and
  apparatus for monitoring vital signs remotely,'' Mar.~27 2008, uS Patent App.
  11/849,051.

\bibitem{zimlichman2012early}
E.~Zimlichman, M.~Szyper-Kravitz, Z.~Shinar, T.~Klap, S.~Levkovich,
  A.~Unterman, R.~Rozenblum, J.~M. Rothschild, H.~Amital, and Y.~Shoenfeld,
  ``Early recognition of acutely deteriorating patients in non-intensive care
  units: Assessment of an innovative monitoring technology,'' \emph{Journal of
  Hospital Medicine}, vol.~7, no.~8, pp. 628--633, 2012.

\bibitem{helfand2016technology}
M.~Helfand, V.~Christensen, and J.~Anderson, ``Technology assessment: Early
  {S}ense for monitoring vital signs in hospitalized patients,'' 2016.

\bibitem{su2018monitoring}
B.~Y. Su, M.~Enayati, K.~Ho, M.~Skubic, L.~Despins, J.~M. Keller, M.~Popescu,
  G.~Guidoboni, and M.~Rantz, ``Monitoring the relative blood pressure using a
  hydraulic bed sensor system,'' \emph{IEEE Transactions on Biomedical
  Engineering}, 2018.

\bibitem{javaid2016elucidating}
A.~Q. Javaid, H.~Ashouri, S.~Tridandapani, and O.~T. Inan, ``Elucidating the
  hemodynamic origin of ballistocardiographic forces: Toward improved
  monitoring of cardiovascular health at home,'' \emph{IEEE Journal of
  Translational Engineering in Health and Medicine}, vol.~4, pp. 1--8, 2016.

\bibitem{rantz2015enhanced}
M.~Rantz, K.~Lane, L.~J. Phillips, L.~A. Despins, C.~Galambos, G.~L. Alexander,
  R.~J. Koopman, L.~Hicks, M.~Skubic, and S.~J. Miller, ``Enhanced registered
  nurse care coordination with sensor technology: Impact on length of stay and
  cost in aging in place housing,'' \emph{Nursing Outlook}, vol.~63, no.~6, pp.
  650--655, 2015.

\bibitem{rantz2015new}
M.~J. Rantz, M.~Skubic, M.~Popescu, C.~Galambos, R.~J. Koopman, G.~L.
  Alexander, L.~J. Phillips, K.~Musterman, J.~Back, and S.~J. Miller, ``A new
  paradigm of technology-enabled 'vital signs' for early detection of health
  change for older adults,'' \emph{Gerontology}, vol.~61, no.~3, pp. 281--290,
  2015.

\bibitem{Noordergraaf1963}
A.~Noordergraaf, P.~D. Verdouw, and H.~B. Boom, ``The use of an analog computer
  in a circulation model,'' \emph{Progress in Cardiovascular Diseases}, vol.~5,
  no.~5, pp. 419--439, 1963.

\bibitem{Wiard2009}
R.~M. Wiard, H.~J. Kim, C.~A. Figueroa, G.~T. Kovacs, C.~A. Taylor, and
  L.~Giovangrandi, ``Estimation of central aortic forces in the
  ballistocardiogram under rest and exercise conditions,'' in \emph{Engineering
  in Medicine and Biology Society, 2009. EMBC 2009. Annual International
  Conference of the IEEE}.\hskip 1em plus 0.5em minus 0.4em\relax IEEE, 2009,
  pp. 2831--2834.

\bibitem{Kim2016}
C.-S. Kim, S.~L. Ober, M.~S. McMurtry, B.~A. Finegan, O.~T. Inan, R.~Mukkamala,
  and J.-O. Hahn, ``Ballistocardiogram: Mechanism and potential for unobtrusive
  cardiovascular health monitoring,'' \emph{Scientific Reports}, vol.~6, p.
  31297, 2016.

\bibitem{bogaard2009right}
H.~J. Bogaard, K.~Abe, A.~V. Noordegraaf, and N.~F. Voelkel, ``The right
  ventricle under pressure: cellular and molecular mechanisms of right-heart
  failure in pulmonary hypertension,'' \emph{Chest}, vol. 135, no.~3, pp.
  794--804, 2009.

\bibitem{di2016deep}
M.~Di~Nisio, N.~van Es, and H.~R. B{\"u}ller, ``Deep vein thrombosis and
  pulmonary embolism,'' \emph{The Lancet}, vol. 388, no. 10063, pp. 3060--3073,
  2016.

\bibitem{jolly2018pulmonary}
M.~Jolly and J.~Phillips, ``Pulmonary embolism: Current role of catheter
  treatment options and operative thrombectomy,'' \emph{Surgical Clinics},
  vol.~98, no.~2, pp. 279--292, 2018.

\bibitem{kerckhoffs2007coupling}
R.~C. Kerckhoffs, M.~L. Neal, Q.~Gu, J.~B. Bassingthwaighte, J.~H. Omens, and
  A.~D. McCulloch, ``Coupling of a 3d finite element model of cardiac
  ventricular mechanics to lumped systems models of the systemic and pulmonic
  circulation,'' \emph{Annals of Biomedical Engineering}, vol.~35, no.~1, pp.
  1--18, 2007.

\bibitem{smith2004minimal}
B.~W. Smith, J.~G. Chase, R.~I. Nokes, G.~M. Shaw, and G.~Wake, ``Minimal
  haemodynamic system model including ventricular interaction and valve
  dynamics,'' \emph{Medical Engineering \& Physics}, vol.~26, no.~2, pp.
  131--139, 2004.

\bibitem{vzavcek1996numerical}
M.~{\v{Z}}{\'a}{\v{c}}ek and E.~Krause, ``Numerical simulation of the blood
  flow in the human cardiovascular system,'' \emph{Journal of Biomechanics},
  vol.~29, no.~1, pp. 13--20, 1996.

\bibitem{liang2005closed}
F.~Liang and H.~Liu, ``A closed-loop lumped parameter computational model for
  human cardiovascular system,'' \emph{JSME International Journal Series C
  Mechanical Systems, Machine Elements and Manufacturing}, vol.~48, no.~4, pp.
  484--493, 2005.

\bibitem{olansen2000closed}
J.~B. Olansen, J.~Clark, D.~Khoury, F.~Ghorbel, and A.~Bidani, ``A closed-loop
  model of the canine cardiovascular system that includes ventricular
  interaction,'' \emph{Computers and Biomedical Research}, vol.~33, no.~4, pp.
  260--295, 2000.

\bibitem{blanco2013dimensionally}
P.~Blanco and R.~Feij{\'o}o, ``A dimensionally-heterogeneous closed-loop model
  for the cardiovascular system and its applications,'' \emph{Medical
  Engineering \& Physics}, vol.~35, no.~5, pp. 652--667, 2013.

\bibitem{hirschvogel2017monolithic}
M.~Hirschvogel, M.~Bassilious, L.~Jagschies, S.~M. Wildhirt, and M.~W. Gee, ``A
  monolithic 3d-0d coupled closed-loop model of the heart and the vascular
  system: Experiment-based parameter estimation for patient-specific cardiac
  mechanics,'' \emph{International Journal for Numerical Methods in Biomedical
  Engineering}, vol.~33, no.~8, p. e2842, 2017.

\bibitem{lakin2003whole}
W.~D. Lakin, S.~A. Stevens, B.~I. Tranmer, and P.~L. Penar, ``A whole-body
  mathematical model for intracranial pressure dynamics,'' \emph{Journal of
  Mathematical Biology}, vol.~46, no.~4, pp. 347--383, 2003.

\bibitem{fritzson2005openmodelica}
P.~Fritzson, P.~Aronsson, H.~Lundvall, K.~Nystr{\"o}m, A.~Pop, L.~Saldamli, and
  D.~Broman, ``The {O}pen{M}odelica modeling, simulation, and development
  environment,'' in \emph{46th Conference on Simulation and Modelling of the
  Scandinavian Simulation Society (SIMS2005), Trondheim, Norway, October 13-14,
  2005}, 2005.

\bibitem{petzold1982dassl}
L.~R. Petzold, ``Description of dassl: a differential/algebraic system
  solver,'' Sandia National Labs., Livermore, CA (USA), Tech. Rep., 1982.

\bibitem{matlab}
I.~MathWorks, \emph{MATLAB: Application program interface guide}.\hskip 1em
  plus 0.5em minus 0.4em\relax MathWorks, 1996, vol.~5.

\bibitem{Accelerometer_Datasheet}
\BIBentryALTinterwordspacing
K.~Inc., ``Kxr94-2283,'' 2014. [Online]. Available:
  \url{http://kionixfs.kionix.com/en/datasheet/KXR94-2283}
\BIBentrySTDinterwordspacing

\bibitem{ADI}
\BIBentryALTinterwordspacing
------, ``Adinstruments powerlab 16/35.'' [Online]. Available:
  \url{https://www.adinstruments.com/products/powerlab}
\BIBentrySTDinterwordspacing

\bibitem{klabunde2011cardiovascular}
R.~Klabunde, \emph{Cardiovascular physiology concepts}.\hskip 1em plus 0.5em
  minus 0.4em\relax Lippincott Williams \& Wilkins, 2011.

\bibitem{Maceira2006Left}
A.~Maceira, S.~Prasad, M.~Khan, and D.~Pennell, ``Normalized left ventricular
  systolic and diastolic function by steady state free precession
  cardiovascular magnetic resonance,'' \emph{Journal of Cardiovascular Magnetic
  Resonance}, vol.~8, no.~3, pp. 417--426, 2006.

\bibitem{Maceira2006Right}
A.~M. Maceira, S.~K. Prasad, M.~Khan, and D.~J. Pennell, ``Reference right
  ventricular systolic and diastolic function normalized to age, gender and
  body surface area from steady-state free precession cardiovascular magnetic
  resonance,'' \emph{European Heart Journal}, vol.~27, no.~23, pp. 2879--2888,
  2006.

\bibitem{Chart}
E.~Lifesciences, ``Normal hemodynamic parameters and laboratory values,''
  \emph{Retrieved on April}, vol.~9, 2014.

\bibitem{mcdonaldblood}
D.~McDonald, ``Blood flow in arteries. 1974,'' \emph{Edward Arnold, London},
  pp. 92--95.

\bibitem{davies2012attenuation}
J.~E. Davies, J.~Alastruey, D.~P. Francis, N.~Hadjiloizou, Z.~I. Whinnett,
  C.~H. Manisty, J.~Aguado-Sierra, K.~Willson, R.~A. Foale, I.~S. Malik
  \emph{et~al.}, ``Attenuation of wave reflection by wave entrapment creates a
  'horizon effect' in the human aorta,'' \emph{Hypertension}, vol.~60, no.~3,
  pp. 778--785, 2012.

\bibitem{davies2007importance}
J.~E. Davies, N.~Hadjiloizou, D.~Leibovich, A.~Malaweera, J.~Alastruey-Arimon,
  Z.~I. Whinnett, C.~H. Manisty, D.~P. Francis, J.~Aguado-Sierra, R.~A. Foale
  \emph{et~al.}, ``Importance of the aortic reservoir in determining the shape
  of the arterial pressure waveform--the forgotten lessons of frank,''
  \emph{Artery Research}, vol.~1, no.~2, pp. 40--45, 2007.

\bibitem{parker2009introduction}
K.~H. Parker, ``An introduction to wave intensity analysis,'' \emph{Medical \&
  Biological Engineering \& Computing}, vol.~47, no.~2, p. 175, 2009.

\bibitem{epstein2015reducing}
S.~Epstein, M.~Willemet, P.~J. Chowienczyk, and J.~Alastruey, ``Reducing the
  number of parameters in 1d arterial blood flow modeling: less is more for
  patient-specific simulations,'' \emph{American Journal of Physiology-Heart
  and Circulatory Physiology}, vol. 309, no.~1, pp. H222--H234, 2015.

\bibitem{noordergraaf1958genesis}
A.~Noordergraaf and C.~E. Heynekamp, ``Genesis of displacement of the human
  longitudinal ballistocardiogram from the changing blood distribution,''
  \emph{American Journal of Cardiology}, vol.~2, no.~6, pp. 748--756, 1958.

\bibitem{cattermole2017normal}
G.~N. Cattermole, P.~M. Leung, G.~Y. Ho, P.~W. Lau, C.~P. Chan, S.~S. Chan,
  B.~E. Smith, C.~A. Graham, and T.~H. Rainer, ``The normal ranges of
  cardiovascular parameters measured using the ultrasonic cardiac output
  monitor,'' \emph{Physiological Reports}, vol.~5, no.~6, 2017.

\bibitem{schwartzenberg2012effects}
S.~Schwartzenberg, M.~M. Redfield, A.~M. From, P.~Sorajja, R.~A. Nishimura, and
  B.~A. Borlaug, ``Effects of vasodilation in heart failure with preserved or
  reduced ejection fraction: implications of distinct pathophysiologies on
  response to therapy,'' \emph{Journal of the American College of Cardiology},
  vol.~59, no.~5, pp. 442--451, 2012.

\bibitem{giovangrandi2012preliminary}
L.~Giovangrandi, O.~T. Inan, D.~Banerjee, and G.~T. Kovacs, ``Preliminary
  results from {B}{C}{G} and {E}{C}{G} measurements in the heart failure
  clinic,'' in \emph{Engineering in Medicine and Biology Society (EMBC), 2012
  Annual International Conference of the IEEE}.\hskip 1em plus 0.5em minus
  0.4em\relax IEEE, 2012, pp. 3780--3783.

\bibitem{bruser2013automatic}
C.~Bruser, J.~Diesel, M.~D. Zink, S.~Winter, P.~Schauerte, and S.~Leonhardt,
  ``Automatic detection of atrial fibrillation in cardiac vibration signals,''
  \emph{IEEE Journal of Biomedical and health informatics}, vol.~17, no.~1, pp.
  162--171, 2013.

\bibitem{etemadi2014tracking}
M.~Etemadi, S.~Hersek, J.~M. Tseng, N.~Rabbani, J.~A. Heller, S.~Roy, L.~Klein,
  and O.~T. Inan, ``Tracking clinical status for heart failure patients using
  ballistocardiography and electrocardiography signal features,'' in
  \emph{Engineering in Medicine and Biology Society (EMBC), 2014 36th Annual
  International Conference of the IEEE}.\hskip 1em plus 0.5em minus 0.4em\relax
  IEEE, 2014, pp. 5188--5191.

\bibitem{reymond2009validation}
P.~Reymond, F.~Merenda, F.~Perren, D.~Rufenacht, and N.~Stergiopulos,
  ``Validation of a one-dimensional model of the systemic arterial tree,''
  \emph{American Journal of Physiology-Heart and Circulatory Physiology}, vol.
  297, no.~1, pp. H208--H222, 2009.

\bibitem{szopos2016mathematical}
M.~Szopos, S.~Cassani, G.~Guidoboni, C.~Prud'Homme, R.~Sacco, B.~Siesky, and
  A.~Harris, ``Mathematical modeling of aqueous humor flow and intraocular
  pressure under uncertainty: towards individualized glaucoma management,''
  \emph{Journal for Modeling in Ophthalmology}, vol.~1, no.~2, pp. 29--39,
  2016.

\bibitem{stone2013unobtrusive}
E.~E. Stone and M.~Skubic, ``Unobtrusive, continuous, in-home gait measurement
  using the microsoft kinect,'' \emph{IEEE Transactions on Biomedical
  Engineering}, vol.~60, no.~10, pp. 2925--2932, 2013.

\bibitem{stone2015average}
E.~Stone, M.~Skubic, M.~Rantz, C.~Abbott, and S.~Miller, ``Average in-home gait
  speed: Investigation of a new metric for mobility and fall risk assessment of
  elders,'' \emph{Gait \& Posture}, vol.~41, no.~1, pp. 57--62, 2015.

\bibitem{skubic2015automated}
M.~Skubic, R.~D. Guevara, and M.~Rantz, ``Automated health alerts using in-home
  sensor data for embedded health assessment,'' \emph{IEEE Journal of
  Translational Engineering in Health and Medicine}, vol.~3, pp. 1--11, 2015.

\bibitem{lydon2015robust}
K.~Lydon, B.~Y. Su, L.~Rosales, M.~Enayati, K.~Ho, M.~Rantz, and M.~Skubic,
  ``Robust heartbeat detection from in-home ballistocardiogram signals of older
  adults using a bed sensor,'' in \emph{Engineering in Medicine and Biology
  Society (EMBC), 2015 37th Annual International Conference of the IEEE}.\hskip
  1em plus 0.5em minus 0.4em\relax IEEE, 2015, pp. 7175--7179.

\bibitem{Enayati2018}
J.~M. K. M.~P. M.~Enayati, M.~Skubic and N.~Z. Farahani, \emph{Sleep Posture
  Classification Using Bed Sensor Data and Neural Networks}, in Engineering in
  Medicine and Biology Society (EMBC), 2018 40th Annual International
  Conference of the IEEE,, 2018.

\bibitem{yang2016sleep}
J.~Yang, J.~M. Keller, M.~Popescu, and M.~Skubic, ``Sleep stage recognition
  using respiration signal,'' in \emph{Engineering in Medicine and Biology
  Society (EMBC), 2016 IEEE 38th Annual International Conference of the}.\hskip
  1em plus 0.5em minus 0.4em\relax IEEE, 2016, pp. 2843--2846.

\bibitem{Avanzolini1988}
G.~Avanzolini, P.~Barbini, A.~Cappello, and G.~Cevenini, ``Cadcs simulation of
  the closed-loop cardiovascular system,'' \emph{International Journal of
  Biomedical Computing}, vol.~22, no.~1, pp. 39--49, 1988.

\bibitem{formaggia2010cardiovascular}
L.~Formaggia, A.~Quarteroni, and A.~Veneziani, \emph{Cardiovascular
  Mathematics: Modeling and simulation of the circulatory system}.\hskip 1em
  plus 0.5em minus 0.4em\relax Springer Science \& Business Media, 2010,
  vol.~1.

\bibitem{Canic2006}
S.~{\v{C}}ani{\'c}, J.~Tamba{\v{c}}a, G.~Guidoboni, A.~Mikeli{\'c}, C.~J.
  Hartley, and D.~Rosenstrauch, ``Modeling viscoelastic behavior of arterial
  walls and their interaction with pulsatile blood flow,'' \emph{SIAM Journal
  on Applied Mathematics}, vol.~67, no.~1, pp. 164--193, 2006.

\bibitem{vcanic2006blood}
S.~{\v{C}}ani{\'c}, C.~J. Hartley, D.~Rosenstrauch, J.~Tamba{\v{c}}a,
  G.~Guidoboni, and A.~Mikeli{\'c}, ``Blood flow in compliant arteries: an
  effective viscoelastic reduced model, numerics, and experimental
  validation,'' \emph{Annals of Biomedical Engineering}, vol.~34, no.~4, pp.
  575--592, 2006.

\bibitem{le2013real}
T.~Q. Le, S.~T. Bukkapatnam, and R.~Komanduri, ``Real-time lumped parameter
  modeling of cardiovascular dynamics using electrocardiogram signals: toward
  virtual cardiovascular instruments,'' \emph{IEEE Transactions on Biomedical
  Engineering}, vol.~60, no.~8, pp. 2350--2360, 2013.

\end{thebibliography}

\end{document}